\documentclass[aps,prd,twocolumn,superscriptaddress]{revtex4}
\usepackage{url}
\usepackage{epsfig}
\usepackage{graphicx}
\usepackage{amsmath}
\usepackage{bm}
\usepackage{setspace}
\usepackage{lscape}
\usepackage[hypertex,urlcolor=blue,colorlinks=true,dvipdfmx]{hyperref}

\newcommand{\beq}{\begin{eqnarray}}
\newcommand{\eeq}{\end{eqnarray}}

\begin{document}

\title{Spectrum of heavy baryons in the quark model}

\author{T. Yoshida}
\affiliation{Department of Physics, Tokyo Institute of Technology, Meguro 152-8551, Japan}
\author{E.Hiyama}
\affiliation{Nishina Center for Accelerator-Based Science, RIKEN, Wako 351-0198, Japan}
\affiliation{Department of Physics, Tokyo Institute of Technology, Meguro 152-8551, Japan}
\affiliation{Advanced Science Research Center, Japan Atomic Energy Agency, Tokai, Ibaraki, 319-1195 Japan}
\author{A.Hosaka}
\affiliation{Research Center for Nuclear Physics (RCNP), Osaka University, Ibaraki, Osaka 567-0047, Japan}
\affiliation{Advanced Science Research Center, Japan Atomic Energy Agency, Tokai, Ibaraki, 319-1195 Japan}
\author{M. Oka}
\affiliation{Department of Physics, Tokyo Institute of Technology, Meguro 152-8551, Japan}
\affiliation{Advanced Science Research Center, Japan Atomic Energy Agency, Tokai, Ibaraki, 319-1195 Japan}
\author{K. Sadato\footnote{present address G-search Ltd. Tamachi 108-0022, Japan.}}
\affiliation{Research Center for Nuclear Physics (RCNP), Osaka University, Ibaraki, Osaka 567-0047, Japan}
\date{\today}

\begin{abstract}
Single- and double- heavy baryons are studied in the constituent quark model.
The model Hamiltonian is chosen as a standard one with two exceptions : (1) The color-Coulomb term depend on quark masses, and (2) an antisymmetric $LS$ force is introduced. 
Model parameters are fixed by the strange baryon spectra, $\Lambda$ and $\Sigma$ baryons.
The masses of the observed charmed and bottomed baryons are, then, fairly well reproduced.
Our focus is on the low-lying negative-parity states, in which the heavy baryons show specific excitation modes reflecting the mass differences of heavy and light quarks. By changing quark masses from the SU(3) limit to the strange quark mass, further to the charm and bottom quark masses, 
we demonstrate that the spectra change from the SU(3) symmetry patterns to the heavy quark symmetry ones.
\end{abstract}

\maketitle

\section{Introduction}

Recent hadron physics has been stimulated by observations of exotic hadrons with heavy quarks.  
So-called $X, Y, Z$ mesons contain most likely a hidden heavy quark and antiquark pair, either $\bar cc$
or $\bar bb$. 
In addition, they may contain light quark and antiquark pair, thus forming a multiquark configuration  near the threshold region of open flavor.
For instance, $X(3872), Z_b(10610, 10850)$ are expected to be hadronic molecules of $D \bar D^*$, $B \bar B^*$ or  $B^* \bar B^*$
via strong correlation of quark and antiquark pair \cite{Voloshin:2007dx} \cite{Ohkoda:2011vj}. Furthermore, the recently discovered penta-quarks, $P_c(4380)$ and $P_c(4450)$, by LHCb \cite{Aaij:2015tga} may also have such a structure.

Theoretically, diquark $qq$ correlations may also play an impotant role, leading to the idea of 
compact tetraquarks \cite{Lee:2007tn} \cite{Maiani:2004vq}.
In fact, the diquark correlations have been considered for long time in many different contexts \cite{Selem:2006nd} to explain the mass ordering of light scalar mesons, weak decays of hyperons, missing nucleon resonances, 
novel phase structure of the quark matter and so on.  
In QCD, the correlation densities of the two light quarks were measured, having indicated 
a strong attraction in a so called good diquark pair \cite{Alexandrou:2006cq}.  
In reality, the evidence should be also seen in masses of excited states.  
Charmed baryons with two light quarks may provide a good opportunity 
for such a study. 

A pioneering work was done  some time ago by Copley et al.\cite{Copley:1979wj} in a constituent quark model, and 
later elaborated by Roberts and Pervin \cite{Roberts:2007ni}.  
They studied various excited states of charmed and bottomed baryons by solving three quark systems explicitly.  
Yet a motivation of the present work is to further point out the behavior of various properties 
of heavy baryons
as functions of the heavy quark mass, smoothly interpolating the SU(3) limit of equal quark masses 
and the heavy quark limit.  
Such a study in different flavor regions is useful to systematically 
understand the nature of spectrum, in particular the roles of the two internal motions 
when baryons are regarded as three-body systems of quarks.  
Then the structure information must show up sensitively in various transition amplitudes 
of decays and productions, which can be studied experimentally as planed at J-PARC and FAIR.

Let us start by briefly showing the essential features of the three quark systems 
with one heavy quark ($Q$) of mass $m_Q$ and the two light quarks ($q$) of equal mass $m_q$
using a non-relativistic quark model with a harmonic oscillator potential for confinement \cite{Copley:1979wj}.  

\begin{figure*}[htbp]
  \begin{center}
    \begin{tabular}{c}
      \begin{minipage}{0.95\hsize}
        \begin{center}
          \includegraphics[trim = 10 410 0 0 ,clip, width=12.0cm]{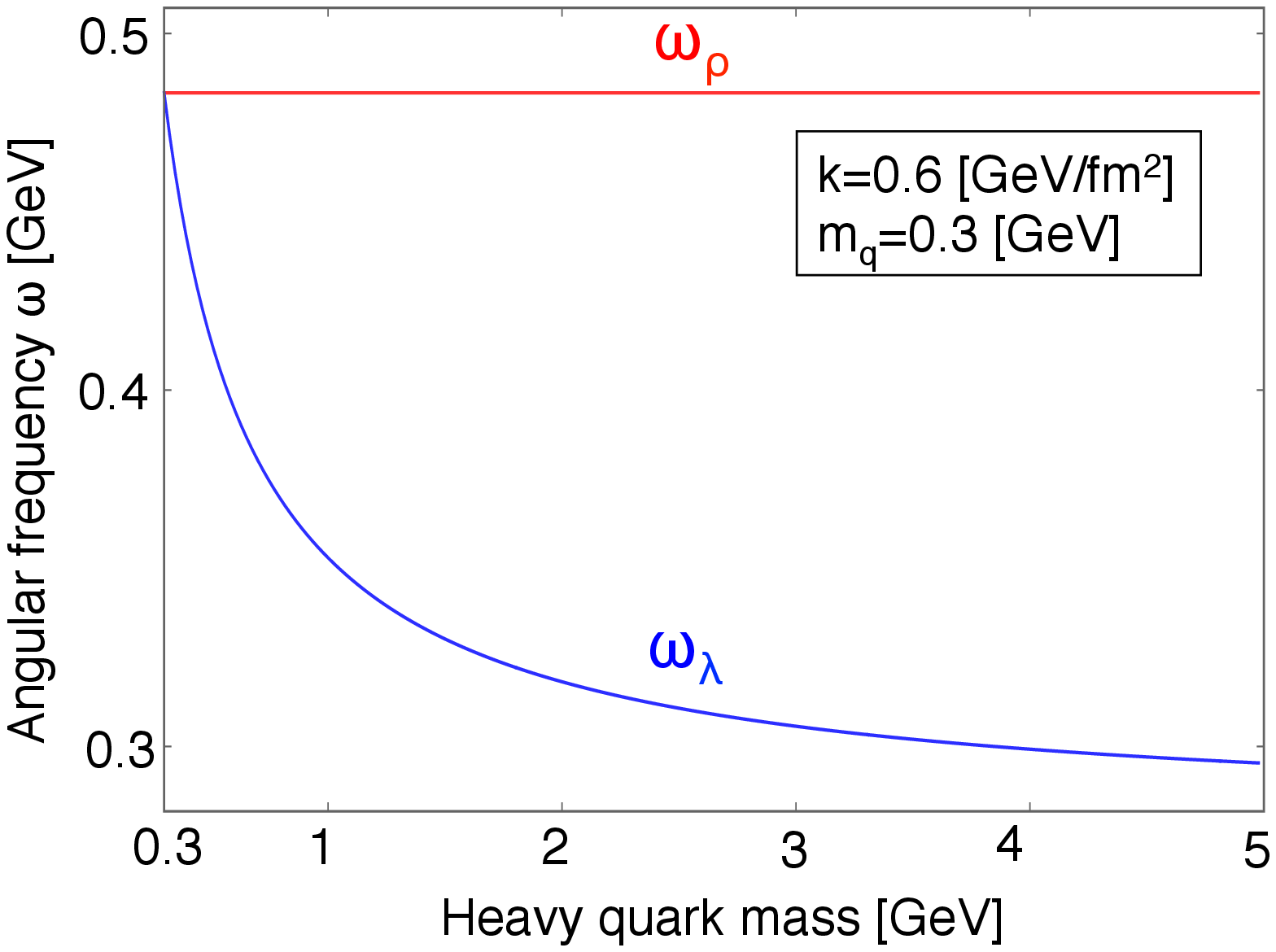}
        \end{center}
     \end{minipage}      
\end{tabular}
    \caption{Heavy quark mass dependence of excited energies of the $\lambda$-mode(red solid line) and the $\rho$-mode(blue solid line) in Eq.(\ref{harmonic})}
    \label{Excited_energy_harmonic}
  \end{center}
\end{figure*}

It is convenient to introduce the Jacobi coordinates, 
$\bm{\lambda}=\bm{r_Q}-\frac{\bm{r_{q1}}+\bm{r_{q2}}}{2}$ and $\bm{\rho}=\bm{r_{q2}}-\bm{r_{q1}}$, 
with obvious notations.  
In the harmonic oscillator potential, the two degrees of freedom decouple and the Hamiltonian can be written 
simply as a sum the two parts,
\begin{eqnarray}
H&=&\sum_i\frac{\bm{p}_i^2}{2m_i}+\sum_{i<j}\frac{k}{2}|\bm{r}_i-\bm{r}_j|^2\nonumber\\
 &=&\frac{\bm{p}_{\rho}^2}{2m_{\rho}}+\frac{\bm{p}_{\lambda}^2}{2m_{\lambda}}+\frac{m_{\rho} {\omega_{\rho}}^2}{2}\bm{\rho}^2+\frac{m_{\lambda}{\omega_{\lambda}}^2}{2}\bm{\lambda}^2,
\label{harmonic}
\end{eqnarray}
where, $m_{\rho}$ and $m_{\lambda}$ denote the reduced masses
\begin{equation}
m_{\rho}=\frac{m_q}{2}\;\;,\;\; m_{\lambda}=\frac{2m_qm_Q}{2m_q+m_Q}.
\end{equation}
and the oscillator frequencies $\omega_{\rho}$ and $\omega_{\lambda}$ are given by
\begin{equation}
\omega_{\rho}=\sqrt{\frac{3k}{2m_{\rho}}} \;\;\;\; \omega_{\lambda}=\sqrt{\frac{2k}{m_{\lambda}}} .
\end{equation}
The ratio of the two excited energy is then given by 
\begin{eqnarray}
\frac{\omega_{\lambda}}{\omega_{\rho}}=\sqrt{\frac{1}{3}(1+2m_q/m_Q)} \leq 1 .
\label{harmo1}
\end{eqnarray}
In the SU(3) limit, equal quark masses, $m_q= m_Q$,  the $\lambda$ and $\rho$ modes degenerate, 
$\omega_{\lambda}=\omega_{\rho}$.
However, when $m_Q > m_q$, the excited energy of the $\lambda$ mode is smaller than that of the $\rho$ mode, $\omega_{\lambda}<\omega_{\rho}$(See Fig.\ref{Excited_energy_harmonic}). 
Thus, we expect that in the heavy quark sector, the $\lambda$ excitation modes become dominant for low lying states of singly heavy quark baryons. 
In contrast, when $m_Q < m_q$,  which corresponds to doubly heavy-quark baryons, we have $\omega_{\lambda}>\omega_{\rho}$, and therefore, the $\rho$ excitation modes become dominant.   
It is shown that this feature is rather general for non-relativistic potential models except for the case when the Coulomb type potential 
of $1/r$ dominates the binding.  

One important symmetry structure realized in the heavy quark hadrons is the heavy quark spin symmetry (HQS) \cite{Yamaguchi:2014era}. 
In the heavy quark limit, the interactions which depend on the spin of the heavy quark disappear. 
Thus, in a single-heavy hadron the heavy quark spin $\bm{s_Q}$ is conserved, i.e., $[H,\bm{s_Q}]=0$, and, with the conservation of the total angular momentum $\bm{J}$, one sees that $\bm{j}\equiv\bm{J}-\bm{s_Q}$, angular momentum carried by the light quarks (including all the orbital angular momenta) is also conserved. We will call $j$ light-spin-component. Consequently, two states whose quantum number are $J=j+1/2$ and $J=j-1/2$ will be degenerate. They form a heavy quark spin doublet, except for $j=0$, which yields HQS singlet. A simple example of HQS doublet is the pair of $\Sigma_Q(1/2^+)$ and $\Sigma_Q(3/2^+)$. The mass differences $\Sigma_s(3/2^+)$$-$$\Sigma_s(1/2^+)$=174 MeV, $\Sigma_c(3/2^+)$$-$$\Sigma_c(1/2^+)$=63 MeV and $\Sigma_b(3/2^+)$$-$$\Sigma_b(1/2^+)$=22 MeV decrease as $m_Q$ becomes large.

We organize this paper as follows.   
In section II, we present our formulation of the non-relativistic constituent quark model.  
The Hamiltonian and the quark interaction are introduced in section II.A;  we  employ a linear potential for quark confinement supplemented by spin-spin, tensor and spin-orbit (LS) forces.  
The anti-symmetric LS force is also needed to guarantee the heavy quark symmetry. 
In II.B, the Gaussian expansion method is introduced to solve the three-quark system.  
When the heavy quark mass is varied from $m_Q=m_q$ to $m_Q\rightarrow\infty$, then the symmetry of the spectrum changes from the SU(3) to the heavy quark spin symmetry.
In section II.C, the relation of the two symmetry limits and mixings of the two internal excitation modes are discussed.  
The results of the present work are presented in section III.  
The results of single-heavy baryons and those of double-heavy baryons are discussed in III.A and III.D, respectively.  
The properties of the $\lambda$ and $\rho$ modes are discussed in III.C in detail.  
In III.D, the heavy quark limit is investigated.  
Finally, a summary is given in Section IV.    

\section{Formalism}

\subsection{Hamiltonian}
In this subsection, we discuss our model Hamiltonian in detail. 
In the non-relativistic quark model, baryons are formed by three valence (constituent) quarks.
They are confined by a confining potential and interact with each other by residual two-body interactions.  
Their internal motions are then described by the two spatial variables  $\bm{\rho}$ and $\bm{\lambda}$. 
In other models of baryons, non-quark degrees of freedom are considered such as constituent gluons and confining fields.  
Their signals in baryon excitations are, however, not yet confirmed in experiments, and are expected to lie at higher energies than the low lying quark excitation modes.  
Empirically these justify the applicability of the quark model, especially for low lying excitation modes.  

Thus our Hamiltonian is written as
\begin{eqnarray}
  H=K+V_{\rm{con}}+V_{\rm{short}}, \label{H} 
\end{eqnarray}
where  the kinetic energy, $K$, the confinement potential, $V_{\rm{con}}$, and the short range interaction, $V_{\rm{short}}$,  are given as
\begin{equation}
 K=\sum_i\left(m_i+\frac{{\bm{p}^2}_i}{2m_i}\right)-K_G \;\; ,\label{Kine}
\end{equation} 
\begin{eqnarray}
V_{\rm{con}}=\sum_{i<j}\frac{br_{ij}}{2} +C \label{Cen}
\end{eqnarray}
\begin{eqnarray}
V_{\rm{short}}&=&\sum_{i<j}\left[-\frac{2\alpha^{\rm{Coul}}}{3r_{ij}}+\frac{{16\pi\alpha}^{\rm{ss}}}{9m_im_j}\bm{s}_i\cdot\bm{s}_j \frac{{\Lambda}^2}{4\pi r_{ij}}\exp(-\Lambda r_{ij}) \right. \nonumber \\
&+&\frac{{\alpha}^{\rm{so}}(1-\exp(-\Lambda r_{ij}))^2}{3{r_{ij}}^3} \nonumber \\
&\times&\left[(\frac{1}{m_i^2}+\frac{1}{m_j^2}+4\frac{1}{m_im_j})\bm{L}_{ij}\cdot\left(\bm{s}_i+\bm{s}_j\right)\right.\nonumber\\
&+&\left.(\frac{1}{m_i^2}-\frac{1}{m_j^2})\bm{L}_{ij}\cdot\left(\bm{s}_i-\bm{s}_j\right)\right] \nonumber \\
&+& \left. \frac{{2\alpha}^{\rm{ten}}(1-\exp(-\Lambda r_{ij}))^2}{3m_im_j{r_{ij}}^3}\left(\frac{3(\bm{s}_i\cdot\bm{ r_{ij}})(\bm{s}_j\cdot\bm{ r_{ij}})}{{r_{ij}}^2}\right.\right.\nonumber\\
&-&\left.\bm{s}_i\cdot\bm{s}_j\biggr)\right]. \label{hyp}
\end{eqnarray}
In Eq.(\ref{Kine}), $m_i$ is the constituent quark mass of the $i$-th quark, 
and the center of mass energy, $K_G$, is subtracted so that the kinetic energy consists only of the $\rho$ and $\lambda$-kinetic energies.  
In Eq.(\ref{Cen}), we employ the linear confinement potential with the $b$ parameter corresponding to the string tension  and $\bm{r}_{ij}=\bm{r}_i-\bm{r}_j$ is the relative coordinate.  
In Eq.(\ref{hyp}), 
$\bm{L}_{ij}=(\bm{r}_i-\bm{r}_j)\times(m_j\bm{p}_i-m_i\bm{p}_j)/(m_i+m_j)$ is the relative orbital angular momentum and  $\bm
{s}_i (= \sigma_i/2)$ is the spin operators of the $i$-th quark.  
The components of Eq.(\ref{hyp}) are inferred by the one-gluon-exchange (OGE), which requires only one coupling constant common to the four terms.  
Practically, however, they may have different origins other than the OGE, and therefore, we treat the four coupling strengths, $\alpha^{\rm{Coul}}$, $\alpha^{\rm{ss}}$, $\alpha^{\rm{so}}$ and $\alpha^{\rm{ten}}$ as independent parameters for better description of baryon masses.  

In order to guarantee the heavy quark symmetry, we introduce anti-symmetric $LS$ force (ALS).  The terms dependent on the heavy quark spin $\bm{s_Q}$ of the $V_{\rm{SLS}}$ and $V_{\rm{ALS}}$ in a single-heavy baryon are given by
\begin{eqnarray}
V_{\rm{SLS}} &\rightarrow& \sum_{i=1,2}\frac{{\alpha}^{\rm{so}}(1-\exp(-\Lambda r_{iQ}))^2}{3{r_{iQ}^3}}\nonumber\\
&\times&\left(\frac{1}{m^2_i}+\frac{1}{m^2_Q}+\frac{4}{m_im_Q}\right)\bm{L}_{iQ}\cdot\bm{s}_Q \\
V_{\rm{ALS}} &\rightarrow&\sum_{i=1,2}\frac{{\alpha}^{\rm{so}}(1-\exp(-\Lambda r_{iQ}))^2}{3{r_{iQ}^3}}\\
&\times&\left(\frac{1}{m^2_i}-\frac{1}{m^2_Q}\right)\bm{L}_{iQ}\cdot\left(-\bm{s}_Q\right)
\end{eqnarray}
where we choose $i=3$ for the heavy quark. Then by summing the parts from SLS and ALS, the $\bm{L_Q}\cdot\bm{s_Q}$ is always proportional to $1/m_Q$ or higher. Thus the $\bm{s}_Q$ dependence disappear in the $m_Q\rightarrow\infty$ limit, and the heavy quark symmetry is guaranteed.

Recently, it was suggested by a Lattice QCD calculation\cite{Kawanai:2011xb} 
that  the strength, $\alpha^{\rm{Coul}}$, of the color Coulomb force depends significantly on the quark mass. In our study, we therefore assume that $\alpha^{\rm{Coul}}$ for the $i-j$ pair of quarks depends on the reduced mass, $\mu_{ij}=\frac{m_im_j}{m_i+m_j}$, as follows, 
\begin{equation}
 \alpha^{\rm{Coul}}=\frac{K}{\mu_{ij}}. \label{Coul}
\end{equation}

We summarize 10 parameters in the Hamiltonian employed here in TABLE \ref{parameter}. The parameters are determined from experimental data of the strange baryon spectrum (See TABLE \ref{strange}). First, we switch off the LS and tensor force to determine the parameters $C$, $\alpha_{\rm{ss}}$, $m_q$, $m_s$ and $\Lambda$, $K$ from the positive parity state. Then, we determine $\alpha^{\rm{so}}$, $b$ from negative parity states. The details how to determine the parameters are as follows.\\ 
\begin{itemize}
\item The constant term $C$

In the constituent quark models, we can predict mass differences between different states, but the absolute values can not be determined. In our work, we introduce the constant $C$ to reproduce the ground state of $\Lambda$(1115) and we assume that the constant $C$ is independent of the constituent quark mass. Namely, we use the same value for the charmed baryons.\\
\item Spin-spin term

The spin-spin term in the hamiltonian is responsible for the splitting among $\Lambda$, $\Sigma$ and  ${\Sigma}^{\ast}$. This term depends on $\alpha^{\rm{ss}}$, $m_q$, $m_s$ and $\Lambda$. Because we have four parameters for three states to be fitted, we fix $m_q=300$ MeV which is the standard value suggested from the magnetic moment of the baryon in the constituent quark model and then we determine the other parameters to reproduce the masses of $\Lambda$, $\Sigma$, ${\Sigma}^{\ast}$. \\
\item The parameter $K$

In our calculation, we introduce $\alpha^{\rm{Coul}}$ as a quark mass dependent form as  given by Eq (\ref{Coul}). Thus, the Coulomb force can contribute to the mass splitting between the ground states of $\Lambda_s(\Sigma_s)$, $\Xi_{ss}$, $\Omega_{sss}$. This force also contributes to the mass differences between the ground state and the excited states. We determine the parameter $K$ to reproduce  $\Xi(1/2^+)$ and the mass difference between the ground state and the excited states. \\
\item  The linear confinement $b$

Our emphasis in the present study is on the P wave states. The parameters which mainly determine the mass differences are $b$ and $K$. $K$ is determined     
from $\Xi(1/2^+)$ as mentioned above and we determine the parameter $b$ to reproduce the splitting between ground state and P-wave state.  \\
  \item The spin-orbit coupling $\alpha^{\rm{so}}$
  
The strength $\alpha^{\rm{so}}$ of the spin-orbit force may be determined by the splitting of the P-wave baryons, such as $\Lambda(1/2^-)$ and $\Lambda(3/2^-)$. However, we do not use the lowest $\Lambda(1/2^-)$, =$\Lambda(1405)$, because various recent studies on the $\Lambda(1405)$ resonance suggests that this is not simply a pure three-quark state, but rather a $N\bar{K}$ molecular-like state. Therefore, we determine the parameter $\alpha^{\rm{so}}$ to reproduce the splitting between the second $\Lambda(1/2^-)$ and $\Lambda(3/2^-)$, namely $\Lambda(1670)$ and $\Lambda(1690)$. Thus, as expected, $\alpha_{\rm{so}}$ becomes very small, much smaller than $\alpha_{ss}$. If the spin-spin and LS forces come only from the OGE, then their values are not consistent.  However, other sources of quark interactions including the relativistic correlations to the confinement and instanton induced interaction(III) may contribute also the LS interaction shown\cite{Takeuchi:1994ma} that the LS force from OGE and III are opposite. Then the discrepancy between $\alpha^{\rm{ss}}$ and $\alpha^{\rm{so}}$ can be explained.\\
\item The strength $\alpha^{\rm{ten}}$

The tensor force in the hamiltonian contributes mainly to the positive parity $\Sigma(1/2^+)$, $\Sigma(3/2^+)$ and the lowest negative states. It has been known that the tensor force is weak and does not contribute much except for generating mixings of $S=1/2$ and $3/2$ states.
We choose $\alpha^{\rm{ten}}$ equal to $\alpha^{\rm{so}}$ \\
\item Charm and bottom quark mass, $m_c,m_b$

We fit the charm quark mass $m_c$ (bottom quark mass $m_b$) to the ground state of $\Lambda_c$ ($\Lambda_b$). These values contribute to the mass splittings as well as the absolute values, but once we determine the other parameters in the strange sector, $m_c$ and $m_b$ are determined uniquely.
\end{itemize}

From Table \ref{strange}, we find that our results reproduce most of the known strange baryon masses, except for the second $J^P=1/2^{+}$ state and the first $J^P=1/2^{-}$. It is well known that the Roper resonance  $N(1440)$, the second $J^P=1/2^{+}$ state, is lighter than lowest $J^P=1/2^{-}$ state, which is incompatible with the quark model predictions. Similarly, in the strange sector, the Roper-like states $\Lambda(1600)$ and  $\Sigma(1660)$, are predicted at higher masses than experiment. The origin of these discrepancies may reside outside the simple three quark picture of the baryons in the quark model. We therefore omit these states from the fitting in the present analysis.
\begin{table*}[htp]
    \begin{tabular}{ccc}

          \begin{tabular}{cccccccccc}
 \\ \hline
$m_q$ & $m_s$&$m_c$ & $m_b$& $b$ &$K$ &$\alpha^{\rm{ss}}$&$\alpha^{\rm{so}}$(=$\alpha^{\rm{ten}}$)&$C$&$\Lambda$\\ 
$[$MeV]&[MeV]&[MeV]&[MeV]&[GeV$^2$]&[MeV]&&&[MeV]&[fm$^{-1}$]\\ \hline \hline
300&510&1750&5112&0.165&90&1.2&0.077&-1139&3.5 \\ \hline

          \end{tabular}
    \end{tabular}
\caption{Parameters in the Hamiltonian. We determine $m_q$, $m_s$, $b$, $K$, $\alpha^{ss}$ and $\Lambda$  to reproduce strange baryons and $m_c$ and $m_b$ are determined from the ground state of $\Lambda_c$ and $\Lambda_b$.}
      \label{parameter}
\end{table*}

\begin{table*}[htbp]
  \begin{center}
    \begin{tabular}{ccc}

      \begin{minipage}{0.3\hsize}
        \begin{center}
          \begin{tabular}{ccc}
(a){\large${\Lambda}_s$} &&\\ \\ \hline
$J^P$ & Theory & Exp. \\ 
 &[MeV] & [MeV] \\ \hline \hline
${\frac{1}{2}}^+$  &1116& 1116 \\
                            & 1799 & 1560-1700\\
                             &1922&1750-1850\\  \hline
 ${\frac{3}{2}}^+$  &1882&1850-1910 \\
                            &2030&\\
                             &2100&\\   \hline
${\frac{5}{2}}^+$  &1891&1815-1825\\
     			& 2045& 2090-2140 \\
			&2143& \\  \hline
${\frac{1}{2}}^-$  &1526 &1405\\ 
                           &1665&1660-1680 \\
                           &1777  &1720-1850\\  \hline
${\frac{3}{2}}^-$  &1537 &1520 \\
                           &1685&1685-1695\\
                           &1810& \\ \hline
${\frac{5}{2}}^-$ &1814&1810-1830  \\ 
                          &2394& \\
                           &2448 &\\ \hline 
          \end{tabular}
        \end{center}
      \end{minipage}
      \begin{minipage}{0.3\hsize}
        \begin{center}
          \begin{tabular}{ccc}
(b){\large${\Sigma}_s$} &&\\ \\ \hline
$J^P$ & Theory & Exp. \\ 
 &[MeV] & [MeV] \\ \hline \hline
${\frac{1}{2}}^+$ & 1197&1192 \\
                            &1895 &1630-1690\\
                            &2016& \\  \hline
${\frac{3}{2}}^+$ & 1391&1385 \\      
                            &2004 &\\
                            &2028 &\\  \hline
${\frac{5}{2}}^+$ &2012&  1900-1935\\
                           &2085 &\\
                           &2091 &\\  \hline
${\frac{1}{2}}^-$ & 1654 & ($\approx$1620)\\
                           &1734 &1730-1800\\ 
                           &1751&  \\  \hline
${\frac{3}{2}}^-$  &1660  & 1665-1685\\
                          &1755& 1900-1950\\
                          &1760&   \\  \hline
${\frac{5}{2}}^-$  &1762 &1770-1780 \\
                          & 2324   &\\
                         & 2427  &\\ \hline 
          \end{tabular}
        \end{center}
      \end{minipage}

      \begin{minipage}{0.3\hsize}
        \begin{center}
          \begin{tabular}{ccc}
(c){\large${\Xi}_{ss}$} &&\\ \\ \hline
$J^P$ & Theory & Exp. \\ 
 &[MeV] & [MeV] \\ \hline \hline
${\frac{1}{2}}^+$  &1325& 1314 \\
                            & 1962 & \\
                             &2131&\\  \hline
 ${\frac{3}{2}}^+$  &1525&1530 \\
                            &2034&\\
                             &2115&\\   \hline
${\frac{5}{2}}^+$  &2040&\\
     			& 2166& \\
			&2211& \\  \hline
${\frac{1}{2}}^-$  &1778 &\\ 
                           &1875& \\
                           &1910 &\\  \hline
${\frac{3}{2}}^-$  &1782 &1820 \\
                           &1877&\\
                           &1920& \\  \hline
${\frac{5}{2}}^-$ &1933& \\ 
                          &2460& \\
                           &2518 &\\ \hline 
          \end{tabular}
        \end{center}
      \end{minipage}
    \end{tabular}
  \end{center}
\caption{Calculated energy spectra and corresponding experimental data of $\Lambda_s$,$\Sigma_s$ and $\Xi_{ss}$. We take the 3-star and 4-star resonances in PDG except for the first $1/2^-$ state of $\Sigma_s$ which has only two stars}
    \label{strange}
\end{table*}

\subsection{Baryon wave-function}

We here consider three quark systems (TABLE.\ref{Flavor}) with one heavy quark, $Q=(c \hbox{ or } b)$, with two or three heavy quarks with the same flavor, i.e., $QQ =(cc \hbox{ or } bb)$ and $QQQ =(ccc \hbox{ or } bbb)$. The remaining quarks are $u$, $d$ or $s$.
We classify the baryons according to the number of heavy quarks, and the strangeness, ${\cal S}$ and the total isospin, $T$.
The last column of the TABLE.\ref{Flavor} shows the isospin wave function where $\eta_0=1$.
\begin{table*}[htbb]
\begin{center}
\begin{tabular}{l||c|c||c}
Heavy baryons&Isospin&Strangeness&isospin wave function\\
&${\cal S}$&$T$\\
\hline 
${\Lambda}_Q=[qq]_{T=0}Q$&0&0& $\left[\left[\eta_{1/2}\eta_{1/2}\right]_{\rm{t}=0}\eta_{0}\right]_{T=0} $  \\
${\Sigma}_Q=[qq]_{T=1}Q$&0&1&$ \left[\left[\eta_{1/2}\eta_{1/2}\right]_{\rm{t}=1}\eta_{1/2}\right]_{T=1} $ \\
${\Xi}_Q=sqQ$&-1& $1/2$ & $ \left[\left[\eta_{0}\eta_{1/2}\right]_{\rm{t}=1/2}\eta_{0}\right]_{T=1/2} $ \\
${\Omega}_{Q}=ssQ$&-2& 0&1\\
\hline
${\Xi}_{QQ} =QQq$&0& 1/2 & $ \left[\left[\eta_{0}\eta_{0}\right]_{\rm{t}=0}\eta_{1/2}\right]_{T=1/2} $ \\ 
${\Omega}_{QQ}=QQs$&-1& $0$&1\\ 
${\Omega}_{QQQ}=QQQ$&0& 0 &1\\
\end{tabular}
\end{center}
\caption{Heavy baryons and their flavor contents. We use the isopin classification so that $q$ stands collectively for the $u$ and $d$ quarks. $Q$ denotes a $c$ or $b$ quark.We do not consider mixing of $c$ and $b$}.
\label{Flavor}
\end{table*}

\begin{figure}
\begin{center}
\includegraphics[trim = 10 620 10 0,clip,width=7cm]{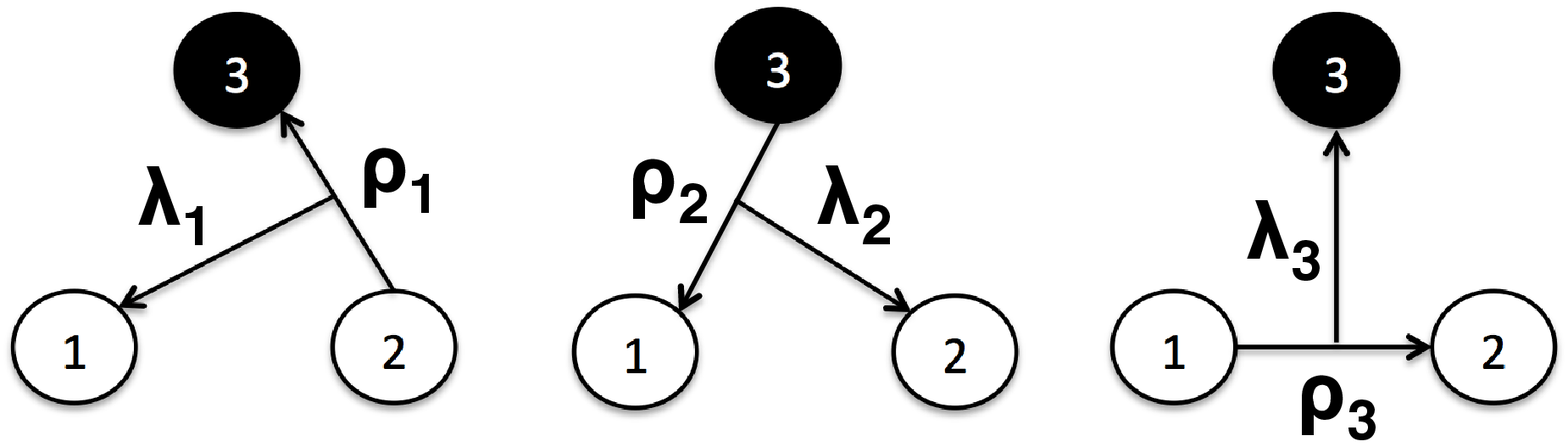}
\end{center}
\caption{Jacobi coordinates for the three body system. 
We place the heavy quark as the 3rd particle in the case of single-heavy baryons, 
while the 1st and 2nd particles  are heavy quarks in double-heavy baryons. }
      \label{jacobi}
\end{figure}

In expressing three-quark wave functions, we introduce three sets of Jacobi coordinates, which we call channels (Fig. \ref{jacobi}).
The Jacobi coordinates in each channel $c$ ($c=1,2,3$) are defined as
\begin{eqnarray}
\bm{\lambda}_c&=&\bm{r_k}-\frac{m_i\bm{r_{i}}+m_j\bm{r_{j}}}{m_i+m_j}, \\
\bm{\rho}_c&=&\bm{r_{j}}-\bm{r_{i}},
\end{eqnarray}
where ($i, j, k$) are given by Table \ref{channel}.
\begin{table}[h]
    \begin{tabular}{cccc}
          \begin{tabular}{cccc}
 \\ \hline
channel&\;\;$i$\;\;&\;\;$j$\;\;&\;\;$k$\;\;\\ \hline \hline
1&2&3&1\\
2&3&1&2\\
3&1&2&3\\  \hline
             \end{tabular}
    \end{tabular}
\caption{The quark assignments ($i$,$j$,$k$) for the Jacobi channels.}
\label{channel}
\end{table}

The total wave function is given as a superposition of the channel wave functions as 
\begin{eqnarray}
\Phi_{\rm{total}}^{JM}=\sum_{c\alpha}C_{c,\alpha}\Phi_{JM,{\alpha}}^{(c)}(\bm{\rho _c},\bm{\lambda_c}),
\label{total-wave-function}
\end{eqnarray}
where the index $\alpha$ represents $\{s,S,\ell, L,I,n,N\}$. Here $s$ is the spin of the ($i, j$) quark pair, $S$ is the total spin,
$\ell$ and $L$ are the orbital angular momentum for the coordinate $\rho$ and $\lambda$, respectively, and $I$ is the total orbital
angular momentum.
The coupling scheme of the spin and angular momenta is as
 \begin{eqnarray}
{\bm s}={\bm s}_i+{\bm s}_j;\; {\bm  s}+{\bm s}_k = {\bm S};\; {\bm \ell} +{\bm L}= {\bm I};\; {\bm S}+{\bm I}={\bm J}.
\end{eqnarray}

The wave function for channel $c$ is given by
\begin{eqnarray}
\Phi_{JM}^{(c)}(\bm{\rho _c},\bm{\lambda_c})=\phi_c\otimes\left[X_{S,s}^{(c)}\otimes \Phi_{\ell,L,\it{I}}^{(c)}\right]_{JM}\otimes H^{(c)}_{T,t}, \label{TotWave}
\end{eqnarray}
where the color wave function, $\phi_c$, the spin wave function, $X_{S}$, the orbital wave function, $\Phi_{I}$, and the isospin wave function, $H_{T}$, are given by
 \begin{equation}
 {\phi}_{\rm{c}}= \frac{1}{\sqrt{6}}(rgb-rbg+gbr-grb+brg-bgr) \label{colorpart}
\end{equation}
 \begin{equation}
 X^{(c)}_{S,s}=\left[\left[\chi_{1/2}(i)\chi_{1/2}(j)\right]_s\chi_{1/2}(k)\right]_{S} \label{spinpart}
\end{equation}
 \begin{equation}
 H^{(c)}_{T.t}=\left[\left[\eta_{\tau_i}(i)\eta_{\tau_j}(j)\right]_t\eta_{\tau_k}(k)\right]_{T} \label{flavorpart}
\end{equation}
 \begin{equation}
 {\Phi}^{(c)}_{\ell,L,\it{I}}=\left[\phi^{(c)}_{\ell}(\bm{\rho_c})\phi^{(c)}_{L}(\bm{\lambda_c})\right]_{\it{I}} \label{spacialpart}
\end{equation}
 \begin{equation}
\phi^{(c)}_{\ell}(\bm{\rho_c})= N_{n\ell}\rho_{c}^{\ell}e^{-\beta_{n}\rho_{c}^2}Y_{\ell m}(\bm{\hat{\rho_c}})\label{spacialpart1}
\end{equation}
 \begin{equation}
\phi^{(c)}_{L}(\bm{\lambda_c})= N_{NL}\lambda_c^Le^{{-\gamma_N
\lambda_c^2}}Y_{LM}(\bm{\hat{\lambda_c}})\label{spacialpart2}.
\end{equation}

In Eq. (\ref{colorpart}), $ r, g ,b$ denote the color of the quark, and the color-singlet wave function is totally anti-symmetric. 
In Eq.(\ref{spinpart}), $\chi_{1/2}$ is the spin wave function of the quark, 
while $\eta_{\tau}$ in Eq.(\ref{flavorpart}) is the isospin wave function with $\tau$ defined by
\begin{equation}
\tau= \left \{
\begin{array}{l}
1/2 \;\;\;\hbox{for $u$, $d$} \\
0 \;\;\;\hbox{for $s$, $c$, $b$}
\end{array}
\right.
\label{eta}
\end{equation}

We consider the quark antisymmetrization for the light quarks, $u$ and $d$, and the heavy quarks, $s$, $c$, $b$, 
separately.
Then for single-heavy baryons, antisymmetrization is applied only to the light quarks.
As the color wave function is always totally anti-symmetric, the spin, isospin and the orbital angular momentum
in the channel (3) should satisfy 
\begin{eqnarray}
&& \ell+s+t=\rm{even} \;\;\; \hbox{for $\Lambda_Q$, $\Sigma_Q$} \label{pau1}
\end{eqnarray}
where $\ell$, $s$, $t$ are the orbital angular momentum, total spin and isospin of the two light quarks. Similarly, the heavy quarks are antisymmetrized in the double-heavy baryons as
\begin{eqnarray}
&& \ell+s+1=\rm{even} \;\;\; \hbox{for $\Xi_{QQ}$, $\Omega_Q$, $\Omega_{QQ}$}\label{pau2}
\end{eqnarray}
where $\ell$, $s$, $t$ are the corresponding ones for the heavy quarks. Considering the antisymmetrization and the combinations of the angular momenta, we obtain
possible assignments of the angular momenta for the low-lying $\Lambda_Q(1/2^+)$ in Table \ref{config}, 
where we take all the combinations satisfying $\ell+L\le 2$.

In solving the Schr\"odinger equation, we use the Gaussian expansion method \cite{Hiyama:2003cu}, where the 
orbital wave functions are expanded, in Eqs. (\ref{spacialpart1}) and (\ref{spacialpart2}), 
by Gaussian functions with the range parameters, $\beta_n$ and $\gamma_N$, 
chosen as:
\begin{eqnarray}
\small{\beta_n}&=&\small{1/r^2_n ,\;r_n=r_1a^{n-1} \;\hbox{($n = 1, \ldots, n_{\rm max}$)},} \\
\small{\gamma_N}&=&\small{1/R^2_N,\;R_N=R_1b^{N-1}\;\hbox{($N = 1, \ldots, N_{\rm max}$)}.}
\end{eqnarray}

\begin{figure}
\begin{center}
\includegraphics[width=8cm]{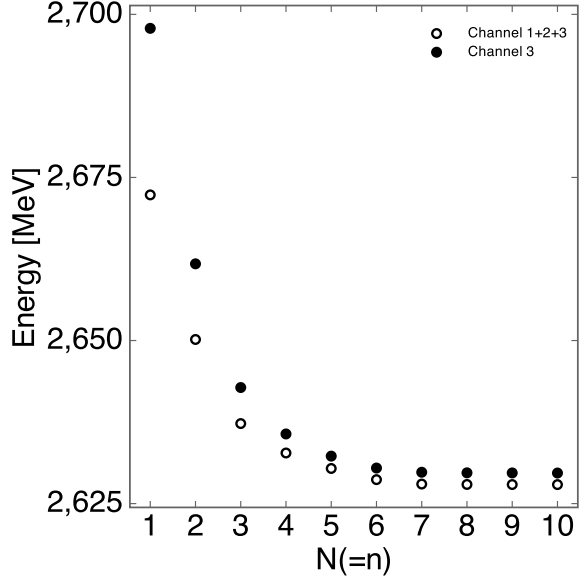}
\end{center}
\caption{Convergence of the energy of the lowest $\Lambda_c(3/2^-)$ 
for increasing the number of bases functions.}
\label{fig}
      \label{Bases}
\end{figure}

In Eqs (\ref{spacialpart1}) ans (\ref{spacialpart2}), $N_{n\ell}$$(N_{NL})$ denotes the normalization constant of the Gaussian basis. The coefficients $C_{c,\alpha}$ of the variational wave function, Eq.(\ref{total-wave-function}), 
are determined by the Rayleigh-Ritz variational principle.
In order to check that the energy converges to the required precision, we change the number of bases and plot the eigen-energy of 
the lowest lying $\Lambda_c(3/2^-)$ in Fig.\ref{Bases}. The filled points are the results from the calculation only using the channel 3,
while the open circles are the results from the three channel calculation (Fig.\ref{jacobi}). 
One sees that when we take only one channel, the convergence is slow and has not yet reached the required precision at 
$N_{\rm max}=n_{\rm max}=10$. 

\begin{table}[h]
    \begin{tabular}{ccc}
          \begin{tabular}{cccccccccc}
 \\ \hline
channel\;\;\;\;&\;$\ell\;\;\;$ & $L$\;\;\;&$I\;\;\;$ & $s$\;\;\;& $S$\;\;\; \\ \hline \hline
3&0&0&0&0&1/2&\\
3&1&1&0&1&1/2&\\
3&1&1&1&1&1/2&\\
3&1&1&1&1&3/2&\\
3&1&1&2&1&3/2&\\  \hline
             \end{tabular}
    \end{tabular}
\caption{Combinations of the spin and orbital angular momenta in channel 3 of the low-lying $\Lambda(1/2^+)$. 
In our study, we restrict the total angular momentum up to 2, $\ell+L=0,2$.}
\label{config}
\end{table}

\subsection{Heavy quark limit}

One of the aims of this paper is to see how the heavy baryon spectrum changes when the heavy quark mass $m_Q$ changes.
Two limits are important: the SU(3) limit with $m_Q=m_q$, and the heavy quark (HQ) limit, $m_Q\to \infty$.

\begin{figure}
\centering
\includegraphics[width=5.5cm]{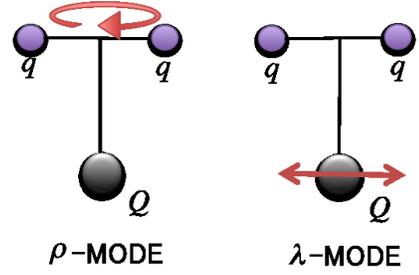}
\caption{The $\rho$- and $\lambda$-modes excitations of  the single-heavy baryon.} \label{figjacobi}
\end{figure}

In the limit $m_Q\rightarrow m_q$, the spectrum is classified by the SU(3) representations.
For instance, the lowest $P$-wave baryons are expected to belong to the SU(6) 70-dimensional representation,
which contains ${}^2\bm{1}$, ${}^2\bm{8}$, ${}^4\bm{8}$, and ${}^2\bm{10}$.
Here the upper index number is the spin multiplicity and the bold number represents the SU(3) multiplicity.
On the other hand, in the HQ limit, $m_Q\to\infty$, as we have discussed in sect.I, the $P$-wave
baryons are better classified by the $\rho$- and $\lambda$-excitation modes (Fig.\ref{figjacobi}).
Here we derive relations between the two pictures.

Let us consider single-heavy $\Lambda_Q$ and $\Sigma_Q$ baryons. We put the heavy quark $Q$ as the 3rd quark.
Then the orbital-spin wave functions of $\Lambda_Q$ and $\Sigma_Q$ in the SU(3) limit are given  by
\begin{eqnarray}
\Psi(\Lambda_Q;{}^2\bm{1})&=&\frac{1}{\sqrt{2}}(X_{1/2,1}\Phi_{1,0,1}-X_{1/2,0}\Phi_{1,1,0}). \label{singletL}\\
\Psi(\Lambda_Q;{}^2\bm{8})&=&\frac{1}{\sqrt{2}}(X_{1/2,1}\Phi_{1,0,1}+X_{1/2,0}\Phi_{1,1,0}) \label{octetL2}\\
\Psi(\Lambda_Q;{}^4\bm{8})&=&X_{3/2,1}\Phi_{1,0,1} \label{octetL1} 
\end{eqnarray}
and
\begin{eqnarray}
 \Psi(\Sigma_Q;{}^2\bm{10})&=&\frac{1}{\sqrt{2}}(X_{1/2,1}\Phi_{1,1,0}+X_{1/2,0}\Phi_{1,0,1})\label{decoupletS1}  \\
\Psi(\Sigma_Q;{}^2\bm{8})&=&\frac{1}{\sqrt{2}}(X_{1/2,1}\Phi_{1,1,0}-X_{1/2,0}\Phi_{1,0,1})\label{octetS2} \\
 \Psi(\Sigma_Q;{}^4\bm{8})&=&X_{3/2,1}\Phi_{1,1,0}.\label{octetS3}
\end{eqnarray}

In the SU(3) limit, the $^2\bm{8}$($S=1/2$) and $^4\bm{8}$($S=3/2$) can be mixed with the spin-spin/spin-orbit forces (If we further argue SU(6), they do not mix). For $m_q < m_Q$, $\Psi(\Lambda_Q;{}^2\bm{1})$ and $\Psi(\Lambda_Q;{}^2\bm{8})$
may mix with each other and in the large $m_Q$ limit, they are reduced to the $\lambda$-mode, $\Phi_{1,1,0}$, and the
$\rho$-mode, $\Phi_{1,0,1}$, excitations. Representing the $\lambda(\rho)$-mode with the total spin $S$ by $^{2S+1}\lambda(^{2S+1}\rho)$, we obtain
\begin{eqnarray}
\Psi(\Lambda_Q;{}^2\lambda)&=&X_{1/2,0}\Phi_{1,1,0}\nonumber\\
&=&\frac{1}{\sqrt{2}}(\Psi(\Lambda_Q;{}^2\bm{8})-\Psi(\Lambda_Q;{}^2\bm{1}))  \\
\Psi(\Lambda_Q;{}^2\rho)&=&X_{1/2,1}\Phi_{1,0,1}\nonumber\\
&=&\frac{1}{\sqrt{2}}(\Psi(\Lambda_Q;{}^2\bm{8})+\Psi(\Lambda_Q;{}^2\bm{1}))  \label{wavefunc1}\\
\Psi(\Lambda_Q;{}^4\rho)&=&X_{3/2,1}\Phi_{1,0,1} =\Psi(\Lambda_Q;{}^4\bm{8}).
\end{eqnarray}
for the $\Lambda_Q$ baryons and
\begin{eqnarray}
\Psi(\Sigma_Q;{}^2\lambda)&=&X_{1/2,1}\Phi_{1,1,0}\nonumber\\
&=&\frac{1}{\sqrt{2}}(\Psi(\Sigma_Q;{}^2\bm{10})+\Psi(\Sigma_Q;{}^2\bm{8}))\\
\Psi(\Sigma_Q;{}^2\rho)&=&X_{1/2,0}\Phi_{1,0,1}\nonumber\\
&=&\frac{1}{\sqrt{2}}(\Psi(\Sigma_Q;{}^2\bm{10})-\Psi(\Sigma_Q;{}^2\bm{8}))\label{wavefunc2}\\
\Psi(\Sigma_Q;{}^4\lambda)&=&X_{3/2,1}\Phi_{1,1,0}=\Psi(\Sigma_Q;{}^4\bm{8}), \label{wavefunc}
\end{eqnarray}
for the $\Sigma_Q$ baryons.

Generally, the $\lambda$-modes appear lower in energy than the $\rho$-modes and they do not mix with each other in the heavy quark limit.  
The two states which are in the same mode but have different spin, ($\Lambda_Q;{}^2\rho$, $\Lambda_Q;{}^4\rho$ and $\Sigma_Q;{}^2\lambda$,
$\Sigma_Q;{}^4\lambda$) may mix even in the heavy quark limit, because the light quark spin-spin force is still alive in this limit. 
For intermediate heavy quark masses, all these states may mix and the wave functions of energy eigenstates show how the mixings change as the heavy quark mass increases.

A similar analysis can be done for other heavy quark baryons. We tabulate, in Table \ref{Lr}, the $\lambda$- and $\rho$-modes 
classification of the $P$-wave heavy quark baryons and their quantum numbers in the Jacobi coordinate channel 3.

In the heavy quark limit, $m_Q\rightarrow\infty$, HQS symmetry becomes exact, where the spin degeneracy of $J=j \pm1/2$ appears.  In this limit, the light component $\bm{j}=\bm{J}-\bm{s}_Q$ and the heavy quark spin $\bm{s_Q}$ are conserved independently, $[H,\bm{s_Q}]=0 \rightarrow[H,\bm{J}-\bm{s_Q}]=[H,\bm{j}]=0$. The basis in which $j$ becomes diagonal can be written in terms of the Jacobi-coordinate basis states Eq.(\ref{TotWave}) for the channel $c=3$ as
\begin{eqnarray}
\Psi(qqQ;j;J)&=& [[\chi_{1/2}(q)\chi_{1/2}(q)]_{s}\Phi_{\ell L I}]_j\chi_{1/2}(Q)]_{J} \nonumber \\
                                                                                                &=&\sum_S(-)^{(s+S+1/2)}\sqrt{(2S+1)(2j+1)}\nonumber \\
                &\times&\left \{
\begin{array}{l}
1/2\hspace{0.3cm}s\hspace{0.3cm}S\\
I\hspace{0.5cm}J\hspace{0.4cm}j\\
\end{array}
\right \}  [X_{S,s}\otimes\Phi_{I\ell L}]_{J}
             \label{HQS_Base}
\end{eqnarray}
 
\begin{table*}[htbb]
\begin{center}

\begin{tabular}{cccccccl}
\hline

\;flavor \;\;&\; $\ell$ \;\;&\; $L$\;\; &\;$I$\;\;& \;$s$ \;\;& \;$S$ \;\;&mode&\;\;\;\;\;\;\;\;\;\;$J$ \\ \hline \hline
& 0 & 1 & 1 & 0 & 1/2 &$^2\lambda$&$1/2^-, 3/2^-$\\ 
${\Lambda}_Q$ &1&0& 1 & 1 & 1/2&$^2\rho$&$1/2^-, 3/2^-$ \\ 
& 1&0& 1 & 1 & 3/2 &$^4\rho$&$1/2^-, 3/2^-, 5/2^-$\\ \hline
& 0 & 1 & 1 & 1 & 1/2 &$^2\lambda$&$1/2^-, 3/2^-$\\ 
${\Sigma}_Q$ &0&1& 1 & 1 & 3/2&$^4\lambda$&$1/2^-,3/2^-, 5/2^-$ \\
& 1&0& 1 & 0 & 1/2 &$^2\rho$&$1/2^-, 3/2^-$\\ \hline
 & 0 & 1 & 1 & 0 & 1/2 &$^2\lambda$&$1/2^-, 3/2^-$\\ 
&1&0& 1 & 1 & 1/2&$^2\rho$&$1/2^-, 3/2^-$ \\ 
${\Xi}_Q$& 1&0& 1 & 1 & 3/2 &$^4\rho$&$1/2^-, 3/2^-, 5/2^-$\\
& 0 & 1 & 1 & 1 & 1/2 &$^2\lambda$&$1/2^-, 3/2^-$\\ 
&0&1& 1 & 1 & 3/2&$^4\lambda$&$1/2^-,3/2^-, 5/2^-$ \\
& 1&0& 1 & 0 & 1/2 &$^2\rho$&$1/2^-, 3/2^-$\\ \hline

& 0 & 1 & 1 & 1 & 1/2 &$^2\lambda$&$1/2^-, 3/2^-$\\ 
${\Xi}_{QQ}$ &0&1& 1 & 1 & 3/2&$^4\lambda$&$1/2^-,3/2^-, 5/2^-$ \\
& 1&0& 1 & 0 & 1/2 &$^2\rho$&$1/2^-, 3/2^-$\\ \hline

& 0 & 1 & 1 & 1 & 1/2 &$^2\lambda$&$1/2^-, 3/2^-$\\ 
${\Omega}_{QQ}$ &0&1& 1 & 1 & 3/2&$^4\lambda$&$1/2^-,3/2^-, 5/2^-$ \\
& 1&0& 1 & 0 & 1/2 &$^2\rho$&$1/2^-, 3/2^-$\\ \hline

${\Omega}_{QQQ}$ & 0 & 1 & 1 & 1 & 1/2 &$^2\lambda$&$1/2^-, 3/2^-$\\ 
& 1&0& 1 & 0 & 1/2 &$^2\rho$&$1/2^-, 3/2^-$\\ \hline
\end{tabular}

\end{center}
\caption{The $\lambda$- and $\rho$-mode assignments of the $P$-wave excitations of ${\Lambda}_Q$,
 ${\Sigma}_Q$, $\Xi_Q$, $\Xi_{QQ}$, $\Omega_{QQ}$ and $\Omega_{QQQ}$. The quantum numbers are given in
 the Jacobi coordinate channel 3.}
\label{Lr}
\end{table*}

\section{Results and Discussion}

\subsection{Energy spectra of single-heavy systems}

We first discuss energy spectra of the single-charmed baryons, $\Lambda_c$, $\Sigma_c$ and $\Omega_c$.
The energies of the charmed baryons are listed in Table \ref{charm} and are illustrated in Fig \ref{FGcharm}.
The mass of the lowest $\Lambda_c$ is used to fix the charm quark mass $m_c$.
The energy differences among the lowest $\Lambda_c(1/2^+)$, $\Sigma_c(1/2^+)$, ${\Sigma_c}^{\ast}(3/2^+)$ are given by $\Sigma_c(1/2^+)$$-$$\Lambda_c(1/2^+)$=175 MeV (exp. 170 MeV), ${\Sigma_c}^{\ast}(3/2^+)$$-$$\Sigma_c(1/2^+)$=71 MeV (exp. 65 MeV), which agree very well to the experimental data.
The mass of the other single-charmed baryons are also well reproduced within 50 MeV deviation.

There are two observed states, $\Lambda_c (2940)$ and $\Sigma_c (2800)$, whose spin and parity have not been assigned. The present calculation indicates that $\Lambda_c (2940)$ can be assigned to one of the following states,  $3/2_{1}^{+}$ (2920MeV), $5/2_{1}^{-}$ (2960MeV),  $1/2_{2}^{-}$ (2890MeV), $1/2_{3}^{-}$ (2933MeV), $3/2_{2}^{-}$ (2917MeV), and $3/2_{3}^{-}$ (2956MeV), while $\Sigma_c (2800)$ may be assigned to one of $1/2_{1}^{-}$ (2802MeV), $3/2_{1}^{-}$ (2807MeV), $1/2_{2}^{-}$ (2826MeV), $3/2_{2}^{-}$ (2837MeV) and $5/2_{1}^{-}$ (2839MeV). Here, $J^P_n$ denotes the $n$-th $J^P$ state. 
Further experimental information, such as decay branching ratios and production rates, will be necessary to determine the quantum numbers of these states.
 
For $S=-2$ baryons, the lowest states of $\Omega_c$($1/2^+$) and $\Omega_c$($3/2^+$) have been experimentally observed. 
We underestimate the mass difference between them by about 20 MeV.

\begin{figure*}[htbp]
  \begin{center}
    \begin{tabular}{c}

      \begin{minipage}{0.998\hsize}
        \begin{center}
   \includegraphics[trim = 35 438 10 0,clip, width=17.6cm]{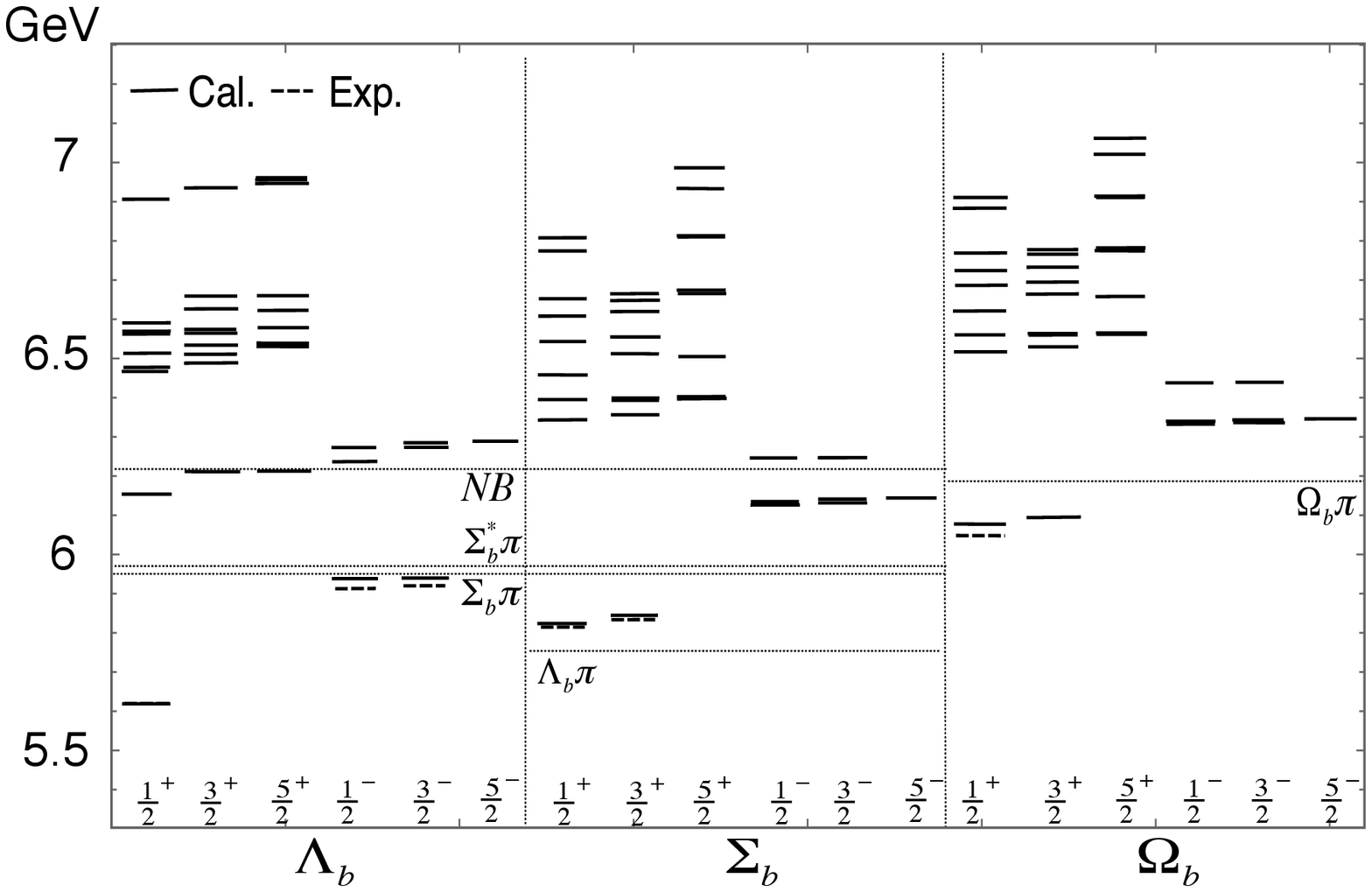}
        \end{center}
      \end{minipage}
    \end{tabular}
    \caption{Calculated energy spectra of $\Lambda_b$, $\Sigma_b$, $\Omega_b$ for $1/2^{+}$, $3/2^{+}$, $5/2^{+}$, $1/2^{-}$, $3/2^{-}$, $5/2^{-}$ (solid line) together with experimental data(dashed line).Several thresholds are also shown by doted line. }
    \label{FGbottom}
  \end{center}
\end{figure*}

The masses of the single-bottom baryons are listed in Table \ref{bottom} and illustrated in Fig \ref{FGbottom}. 
The ground state $\Lambda_b$ is fitted to the experimental data of Particle Data Group . 
The mass differences among $\Lambda_b$, $\Sigma_b$, and ${\Sigma_b}^{\ast}$ are  $\Sigma_b(1/2^+)$$-$$\Lambda_b(1/2^+)$=188MeV, ${\Sigma_b}^{\ast}(3/2^+)$$-$$\Sigma_b(1/2^+)$=21 MeV experimentally, 
while our calculation gives $\Sigma_b(1/2+)$$-$$\Lambda_b(1/2^+)$=195MeV, and ${\Sigma_b}^{\ast}(3/2^+)$$-$$\Sigma_b(1/2^+)$=22 MeV. Thus, we find that the low lying positive-parity states are reproduced  within 10 MeV deviation. 

The negative parity $\Lambda_b$ states, $\Lambda_b$(5912) and $\Lambda_b$(5920), have been discovered recently.
Their mass difference is about 8 MeV in experiment while it is 1 MeV in our prediction.
For $S=-2$ bottom baryons, $\Omega_b$(1/2$^+)$, our estimate of the mass is 6076 MeV, which is higher than the experimental value, 6015 MeV.


\begin{table*}[htbp]
  \begin{center}
    \begin{tabular}{ccc}

      \begin{minipage}{0.3\hsize}
        \begin{center}
          \begin{tabular}{ccc}
(a){\large${\Lambda}_c$} &&\\ \\ \hline
$J^P$ & Theory & Exp. \\ 
 &[MeV] & [MeV] \\ \hline \hline
${\frac{1}{2}}^+$  &2285& 2285 \\
                            & 2857 &         \\
                             &3123&\\  \hline
 ${\frac{3}{2}}^+$  &2920& \\
                            &3175&\\
                             &3191&\\   \hline
${\frac{5}{2}}^+$  &2922&2881\\
     			& 3202&  \\
			&3230 & \\  \hline
${\frac{1}{2}}^-$  &2628 &2595\\ 
                           &2890& \\
                           &2933  &\\  \hline
${\frac{3}{2}}^-$  &2630 &2628 \\
                           &2917&\\
                           &2956& \\  \hline
${\frac{5}{2}}^-$ &2960 &  \\ 
                          &3444& \\
                           &3491 &\\ \hline 
          \end{tabular}
        \end{center}
      \end{minipage}
      \begin{minipage}{0.3\hsize}
        \begin{center}
          \begin{tabular}{ccc}
(b){\large${\Sigma}_c$} &&\\ \\ \hline
$J^P$ & Theory & Exp. \\ 
 &[MeV] & [MeV] \\ \hline \hline
${\frac{1}{2}}^+$ & 2460&2455 \\
                            &3029 &\\
                            &3103& \\  \hline
${\frac{3}{2}}^+$ & 2523  &2518 \\      
                            &3065 &\\
                            &3094 &\\  \hline
${\frac{5}{2}}^+$ &3099&  \\
                           &3114 &\\
                           &3191 &\\  \hline
${\frac{1}{2}}^-$ & 2802 & \\
                           &2826  &\\ 
                           &2909&  \\  \hline
${\frac{3}{2}}^-$  &2807  & \\
                          &2837  & \\
                          &2910&   \\  \hline
${\frac{5}{2}}^-$  &2839 & \\
                          & 3316   &\\
                         & 3521  &\\ \hline 
          \end{tabular}
        \end{center}
      \end{minipage}
      \begin{minipage}{0.3\hsize}
        \begin{center}
          \begin{tabular}{ccc}
(c) {\large${\Omega}_c$} \\  \\ \hline 
$J^P$ & Theory & Exp. \\ 
 &[MeV] & [MeV] \\ \hline \hline
 ${\frac{1}{2}}^+$  &2731 &2698  \\
                             &3227&\\
                             &3292& \\  \hline
 ${\frac{3}{2}}^+$ &2779 &2768\\                             
                            &3257&\\
                            &3285&\\  \hline
${\frac{5}{2}}^+$ &3288&\\                         
                           &3299&\\
                           & 3359&\\   \hline
${\frac{1}{2}}^-$  &3030& \\
                           &3048&\\
                           &3110&\\  \hline
${\frac{3}{2}}^-$  &3033&\\
                           &3056&\\
                           &3111 &\\  \hline
${\frac{5}{2}}^-$  &3057&\\
                           &3477&\\
                           &3620&\\  \hline
          \end{tabular}
        \end{center}
      \end{minipage}

    \end{tabular}
  \end{center}
\caption{Calculated energy spectra and experimental result of $\Lambda_c$, $\Sigma_c$, $\Omega_c$}
\label{charm}
\end{table*}

\begin{table*}[htbp]
  \begin{center}
    \begin{tabular}{cccc}

      \begin{minipage}{0.3\hsize}
        \begin{center}
          \begin{tabular}{ccc}
(a) {\large${\Lambda}_b$} &&\\ \\ \hline
$J^P$ & Theory & Exp. \\ 
 &[MeV] & [MeV] \\ \hline \hline
${\frac{1}{2}}^+$  &5618& 5624 \\
                            & 6153 &         \\
                             &6467&\\  \hline
 ${\frac{3}{2}}^+$  &6211& \\
                            &6488&\\
                             &6511&\\   \hline
${\frac{5}{2}}^+$  &6212& \\
     			& 6530&  \\
			&6539& \\  \hline
${\frac{1}{2}}^-$  & 5938&5912\\ 
                           &6236& \\
                           &6273 &\\  \hline
${\frac{3}{2}}^-$  &5939 &5920\\
                           &6273&\\
                           &6285& \\  \hline
${\frac{5}{2}}^-$ &6289&  \\ 
                          &6739& \\
                           &6786 &\\ \hline 
          \end{tabular}
        \end{center}
      \end{minipage}
      \begin{minipage}{0.3\hsize}
        \begin{center}
          \begin{tabular}{ccc}
(b){\large${\Sigma}_b$} &&\\ \\ \hline
$J^P$ & Theory & Exp. \\ 
 &[MeV] & [MeV] \\ \hline \hline
${\frac{1}{2}}^+$ & 5823 &5815 \\
                            &6343 &\\
                            &6395& \\  \hline
${\frac{3}{2}}^+$ & 5845  &5835 \\      
                            &6356 &\\
                            &6393 &\\  \hline
${\frac{5}{2}}^+$ &6397&  \\
                           &6402 &\\
                           & 6505&\\  \hline
${\frac{1}{2}}^-$ & 6127& \\
                           &6135  &\\ 
                           &6246&  \\  \hline
${\frac{3}{2}}^-$  &6132 & \\
                          &6141& \\
                          &6246&   \\  \hline
${\frac{5}{2}}^-$  &6144 & \\
                          &6592  &\\
                         & 6834    &\\ \hline 
          \end{tabular}
        \end{center}
      \end{minipage}
      \begin{minipage}{0.3\hsize}
        \begin{center}
          \begin{tabular}{ccc}
(c) {\large${\Omega}_b$} \\  \\ \hline 
$J^P$ & Theory & Exp. \\ 
 &[MeV] & [MeV] \\ \hline \hline
 ${\frac{1}{2}}^+$  & 6076&6048  \\
                             &6517&\\
                             &6561& \\  \hline
 ${\frac{3}{2}}^+$ &6094&\\                             
                            &6528&\\
                            &6559&\\  \hline
${\frac{5}{2}}^+$ &6561&\\                         
                           &6566&\\
                           & 6657&\\   \hline
${\frac{1}{2}}^-$  &6333& \\
                           &6340&\\
                           &6437&\\  \hline
${\frac{3}{2}}^-$  &6336&\\
                           &6344&\\
                           &6438 &\\  \hline
${\frac{5}{2}}^-$  &6345&\\
                           &6728&\\
                           &6919&\\  \hline
          \end{tabular}
        \end{center}
      \end{minipage}

    \end{tabular}
  \end{center}
\caption{Calculated energy spectra and experimental result of $\Lambda_b$, $\Sigma_b$, $\Omega_b$}
\label{bottom}
\end{table*}
\begin{table*}[htbp]
  \begin{center}
    \begin{tabular}{ccc}

      \begin{minipage}{0.4\hsize}
        \begin{center}
          \begin{tabular}{ccccc}
(a) {\large$\Xi_{cc}$ } &&\\ \\ \hline
$J^P$ & Theory & Exp.& \cite{Namekawa:2013vu} &\cite{Roberts:2007ni}  \\ 
 &[MeV] & [MeV]&& \\ \hline \hline
 ${\frac{1}{2}}^+$  &3685& 3512&3603$\pm$15$\pm$16&3674 \\
                            & 4079 &         &&4029\\
                             &4159&&&\\  \hline
 ${\frac{3}{2}}^+$  &3754& &3706$\pm$22$\pm$16&3753\\
                            &4114&&&4042\\
                             &4131&&&\\   \hline
${\frac{5}{2}}^+$  &4115&&&4047\\
     			& 4164&  &&4091\\
			&4348 & &&\\  \hline
${\frac{1}{2}}^-$  &3947 &&&3910\\ 
                           &4135& &&4074\\
                           &4149 &&&\\  \hline
${\frac{3}{2}}^-$  &3949 &&& 3921\\
                           &4137&&&4078\\
                           &4159& &&\\  \hline
${\frac{5}{2}}^-$ &4163&  &&4092\\ 
                          &4488& &&\\
                           &4534& &&\\ \hline 
          \end{tabular}
        \end{center}
      \end{minipage}
      \begin{minipage}{0.4\hsize}
        \begin{center}
          \begin{tabular}{ccccc}
(b){\large$\Xi_{bb}$} &&\\ \\ \hline
$J^P$ & Theory &\cite{Roberts:2007ni}  \\ 
 &[MeV] & [MeV]& \\ \hline \hline
 ${\frac{1}{2}}^+$  &10314& 10340 \\
                            & 10571&\\
                             &10612&\\  \hline
 ${\frac{3}{2}}^+$  &10339& 10367\\
                            &10592&\\
                             &10593&\\   \hline
${\frac{5}{2}}^+$  &10592&10676\\
     			& 10613  &\\
			&  10809 &\\  \hline
${\frac{1}{2}}^-$  &10476&10493\\ 
                           &10703&\\
                           & 10740&\\  \hline
${\frac{3}{2}}^-$  &10476&10495 \\
                           &10704&\\
                           &10742&\\  \hline
${\frac{5}{2}}^-$ &10759&10713\\ 
                          &10973& \\
                           &11004&\\ \hline 
          \end{tabular}
        \end{center}
      \end{minipage} 
    \end{tabular}
  \end{center}
\caption{Calculated energy spectra and experimental result of $\Xi_{cc}$ and $\Xi_{bb}$}
\label{gzai}
\end{table*}

\begin{table*}[htbp]
  \begin{center}
    \begin{tabular}{ccc}

      \begin{minipage}{0.4\hsize}
        \begin{center}
          \begin{tabular}{ccccc}
(a) {\large$\Omega_{cc}$ } &&\\ \\ \hline
$J^P$ & Theory & \cite{Namekawa:2013vu}&\cite{Roberts:2007ni}  \\ 
 &[MeV] &  \\ \hline \hline
 ${\frac{1}{2}}^+$  &3832&3704$\pm$5$\pm$16& 3815 \\
                            & 4227 &  &4180\\
                             &4295&&\\  \hline
 ${\frac{3}{2}}^+$  &3883& 3779$\pm$6$\pm$17&3876\\
                            &4263&&4188\\
                             &4265&&\\   \hline
${\frac{5}{2}}^+$  &4264&&4202\\
     			&4299& &4232\\
			& 4410 & &\\  \hline
${\frac{1}{2}}^-$  &4086&&4046\\ 
                           &4199& &4135\\
                           &4210 &&\\  \hline
${\frac{3}{2}}^-$  &4086 &&4052\\
                           &4201&&4140\\
                           &4218& &\\  \hline
${\frac{5}{2}}^-$ &4220& &4152 \\ 
                          &4555& &\\
                           &4600& &\\ \hline 
          \end{tabular}
        \end{center}
      \end{minipage}
      \begin{minipage}{0.4\hsize}
        \begin{center}
          \begin{tabular}{ccccc}
(b){\large$\Omega_{bb}$} &&\\ \\ \hline
$J^P$ & Theory & \cite{Roberts:2007ni}  \\ 
 &[MeV] &  \\ \hline \hline
 ${\frac{1}{2}}^+$  &10447& 10454 \\
                            & 10707& 10693&\\
                             &10744&&\\  \hline
 ${\frac{3}{2}}^+$  &10467& 10486\\
                            &10723&10721&\\
                             &10730&&&\\   \hline
${\frac{5}{2}}^+$  &10729&10720\\
     			&10744& 10734 &\\
			& 10937 & &\\  \hline
${\frac{1}{2}}^-$  &10607 &10616\\ 
                           &10796& 10763&\\
                           &10803 &&\\  \hline
${\frac{3}{2}}^-$  &10608 &10619 \\
                           &10797&10765&\\
                           &10805& &\\  \hline
${\frac{5}{2}}^-$ &10808& 10766\\ 
                          &11028& &\\
                           & 11059& &\\ \hline 
          \end{tabular}
        \end{center}
      \end{minipage} 
    \end{tabular}
  \end{center}
\caption{Calculated energy spectra and experimental result of $\Omega_{cc}$ and $\Omega_{bb}$}
\label{omega}
\end{table*}

\subsection{Energy spectra of double-heavy baryon systems}

\begin{figure*}[htbp]
  \begin{center}
    \begin{tabular}{c}

         \begin{minipage}{0.998\hsize}
        \begin{center}
   \includegraphics[trim = 30 410 0 0,clip, width=16.8cm]{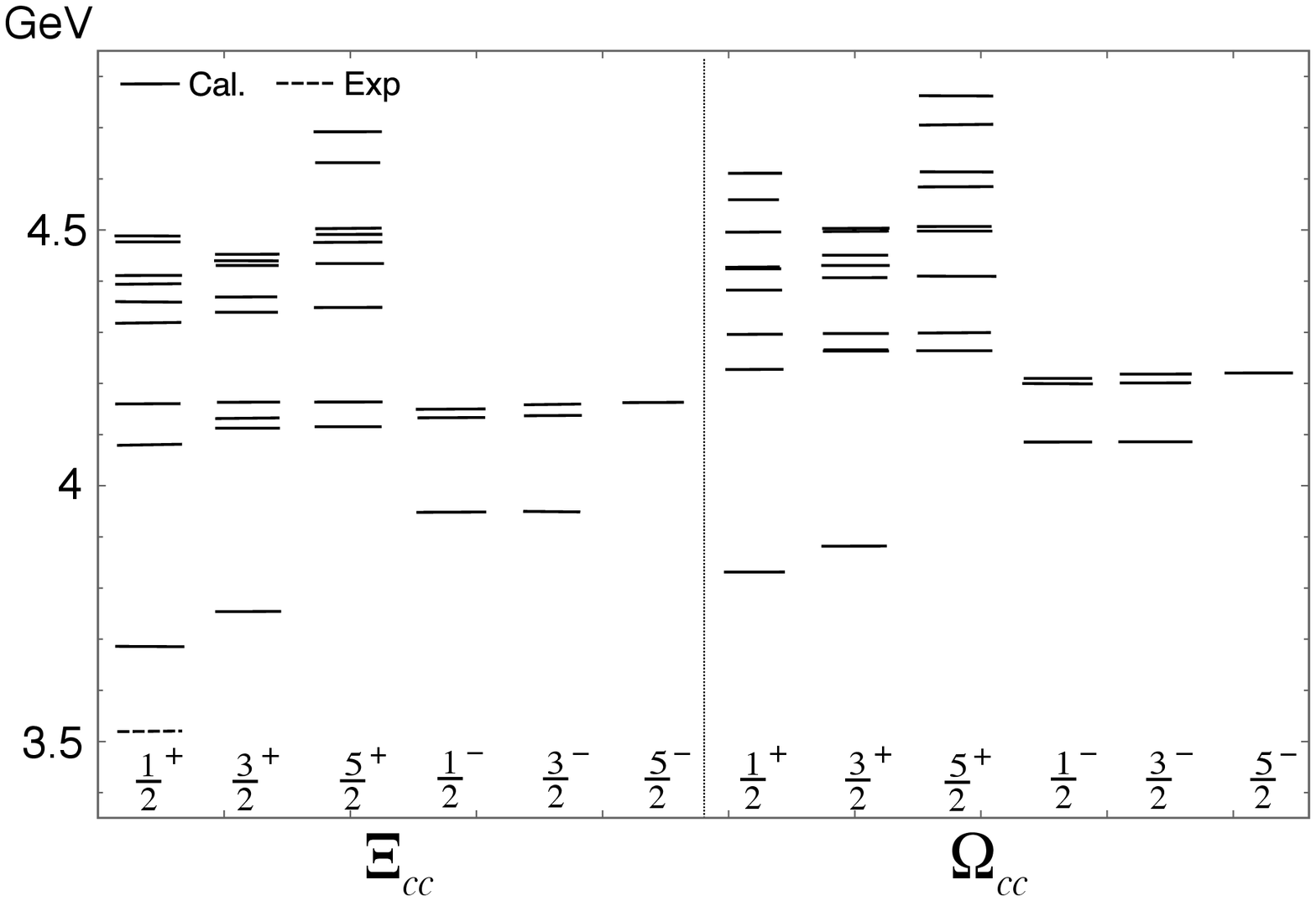}
        \end{center}
      \end{minipage}
    \end{tabular}

    \caption{Calculated energy spectra of $\Xi_{cc}$, $\Omega_{cc}$ for $1/2^{+}$, $3/2^{+}$, $5/2^{+}$, $1/2^{-}$, $3/2^{-}$, $5/2^{-}$(solid line) together with experimental data (dashed line). }
    \label{FGdoublech}
  \end{center}
\end{figure*}

\begin{figure*}[htbp]
  \begin{center}
    \begin{tabular}{c}

         \begin{minipage}{0.998\hsize}
        \begin{center}
   \includegraphics[trim = 30 400 0 0 clip, width=16.8cm]{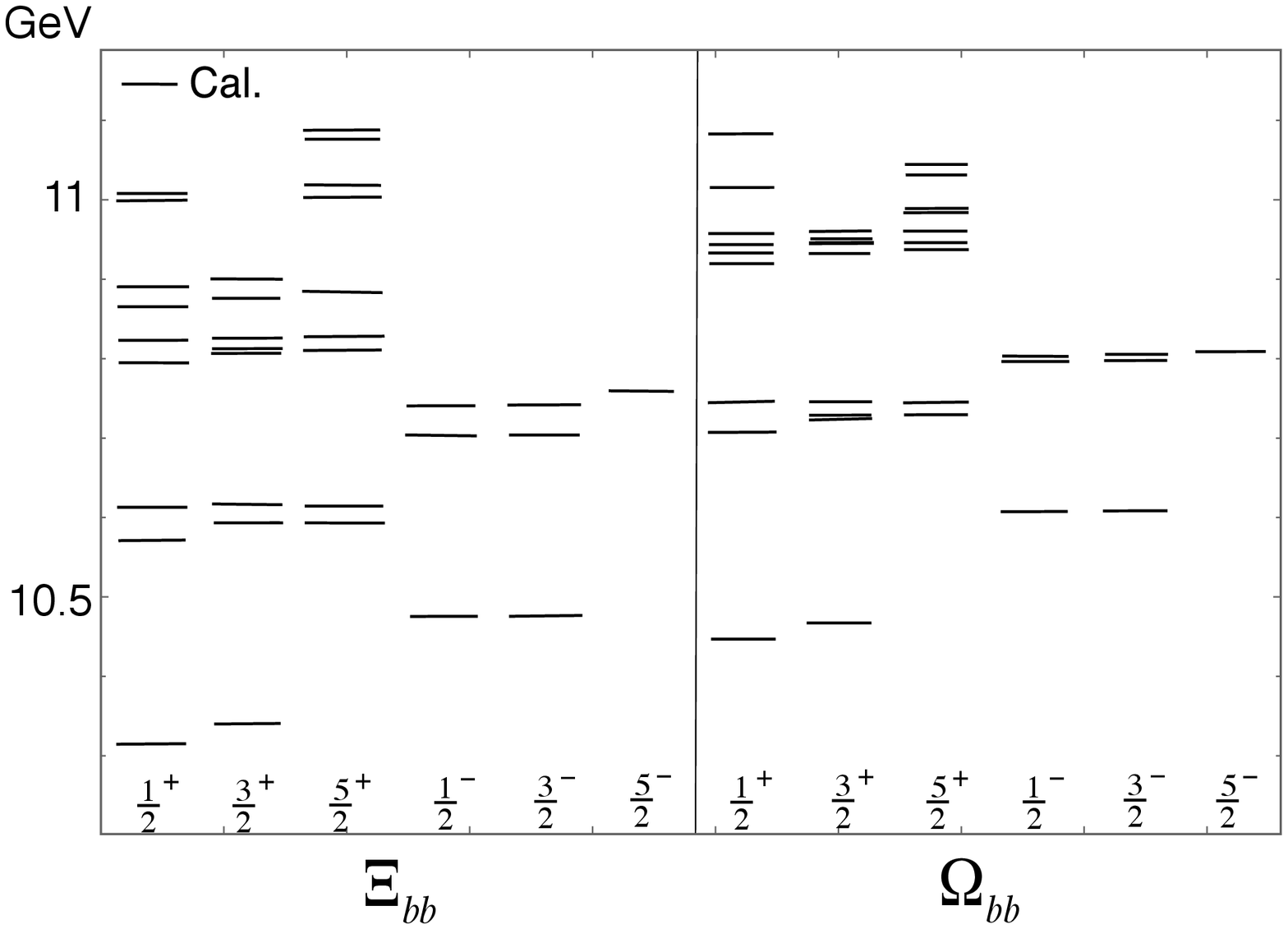}
        \end{center}
      \end{minipage}
    \end{tabular}

    \caption{Calculated energy spectra of $\Xi_{bb}$, $\Omega_{bb}$ for $1/2^{+}$, $3/2^{+}$, $5/2^{+}$, $1/2^{-}$, $3/2^{-}$, $5/2^{-}$ (solid line) together with experimental data (dashed line). }
    \label{FGdoublebt}
  \end{center}
\end{figure*}

TABLE \ref{gzai}, \ref{omega} and Figs.\ref{FGdoublech} and \ref{FGdoublebt} show the calculated energy spectra and experimental data for double-heavy baryons.  
Lattice QCD \cite{Namekawa:2013vu} \cite{Na:2008hz}
 and quark models \cite{Roberts:2007ni} \cite{Albertus:2006ya} predicted the masses of double-heavy baryons and variations among the model calculations are large, compared to those in the single-heavy baryons.

The calculated mass of the lowest $\Xi_{cc}$ state is 3685 MeV, which
 is much higher than the experimental observations 
by SELEX\cite{Moinester:2002uw}, 3519 MeV. 
However, the other experimental searches by BARBAR\cite{Aubert:2006qw}, Belle\cite{Chistov:2006zj} and LHCb\cite{Ogilvy:2013xfa}, could not confirm this state. 
Our prediction is consistent with the recent lattice result as well as the other quark model calculations.
 
We predict that the lowest $\Xi_{bb}$ state is $\Xi_{bb}({\frac{1}{2}}^+)$=10314 MeV followed by $\Xi_{bb}({\frac{3}{2}}^+)$=10339 MeV.

\subsection{$\lambda$ mode and $\rho$ mode structures in heavy baryon systems}

Now we compare the heavy baryon spectra for the strange sector and the heavier sector ($c$ and $b$) and clarify the quark dynamics in the heavy baryon. Strange baryons are conventionally analyzed by the $SU(3)_f$ symmetry. When the strange quark is replaced by a heavier quark, $c$ or $b$, we can study the dynamics of the two light quarks, which may be regarded as a diquark. From this point of view, one sees two distinct excitation modes, $\lambda$ and $\rho$ modes. The $\lambda$-mode
state is composed of the $(qq)_{\ell=0}$ diquark with $\it{L}$=1 excitation relative to the heavy quark, $Q$, while the $\rho$-mode state has an excited diquark $(qq)_{\ell=1}$ in the $L=0$ orbit around $Q$. 

As is discussed in Sec.I, the $\lambda$- and $\rho$-modes are largely mixed in the SU(3) limit in the light quark sector. This mixing is induced mainly by the spin-spin interaction. Because the spin dependent interaction for the heavy quark is weak, the $\lambda$- and $\rho$-modes are well separated for the charm and bottom baryons. Then, each P-wave state is dominated and characterized either by the $\lambda$-mode or $\rho$-mode.

In order to demonstrate these properties quantitatively, we change the heavy quark mass, $m_Q$, from 300 MeV to 6 GeV and analyze the excitation energies and wave functions. Fig. \ref{LAMBDA_SIGMA_MASS_DEP}  shows the spectra of $\Lambda_Q$ and $\Sigma_Q$ as functions of $m_Q$. One sees that the splitting between the 1st and 2nd $1/2^-$ state of $\Lambda_Q$ increases rapidly from 100 MeV in the SU(3) limit to 300 MeV in the heavy quark limit when $m_Q$ increases. This behavior is due to the $\lambda-\rho$ splitting as demonstrated by the harmonic osillator model (in Fig.\ref{Excited_energy_harmonic}). Namely, the lowest state becomes dominated by the $\lambda$-mode as $m_Q$ becomes large. This is confirmed in Fig.\ref{LAMBDA_SIGMA_PROB}, where the  $\lambda$- and $\rho$-mode probabilities of the lowest $1/2^-$ state are plotted as functions of $m_Q$. One sees that the state is almost purely in the $\lambda$ mode at $m_Q \geq1.5$ GeV; the $\lambda$ dominance is seen even at $m_Q=510$ MeV. As is classified in TABLE  \ref{Lr}, the quark model predicts seven P-wave $\Lambda_Q$ excitations, $(1/2^-)^3$, $(3/2^-)^3$, $(5/2^-)$. They split into the $(1/2^-, 3/2^-)$ $\lambda$ modes and $(1/2^-)^2$, $(3/2^-)^2$, $5/2^-$ $\rho$ modes. In Fig. \ref{LAMBDA_SIGMA_MASS_DEP}, one sees clear splitting ($\approx$350 MeV) of two low lying $\lambda$-modes and five higher $\rho$ mode states.

The P-wave $\Sigma_Q$ has also seven states in the quark model, $(1/2^-)^3$, $(3/2^-)^3$, $(5/2^-)$. One sees that they are classified into the $(1/2)^2$, $(3/2^-)^2$, $(5/2^-)$ $\lambda$ modes and $(1/2^-,3/2^-)$ $\rho$ modes from Fig.\ref{LAMBDA_SIGMA_MASS_DEP}. The $\lambda$- and $\rho$- modes are separated more slowly than $\Lambda_Q$ as $m_Q$ increases, and the $\lambda$ dominance is seen at $m_Q\geq1750$ MeV. The difference comes from the interaction between light quarks which forms the diquark. The diquark in $\Sigma_Q$ has spin 1 and the spin-spin interaction is repulsive for the $\lambda$ mode, while the $\rho$ mode has a diquark state of spin 0 and the spin-spin interaction is attractive. Therefore, the difference between the excitation energies of the two modes is small compared to $\Lambda_Q$. Thus, the splitting between the excitation energies of two modes is larger for $\Lambda_Q$ and smaller for $\Sigma_Q$ compared with the case in which there is no spin-spin force as we see in Sec.I.
As a result, the change of the probability of two modes in the $\Sigma_Q$ case is more slow than the $\Lambda_Q$ case as shown in Fig.\ref{LAMBDA_SIGMA_PROB}.

In the case of double-heavy baryon, the $\lambda$-mode state is composed of the $(QQ)_{\ell=0}$ heavy diquark with the light quark $q$, while the $\rho$-mode state has the excited heavy diquark  $(QQ)_{\ell=1}$ in the $L=0$ orbit around $q$.
 The combinations of angular momentum are the same as the $\Sigma_Q$ case which is shown in TABLE \ref{Lr}, but the behavior of $\lambda$- and $\rho$ modes are different because $\Xi_{QQ}$, or $\Omega_{QQ}$ contains heavy diquark. As mentioned in Sec.I,  $\omega_{\lambda}$ is larger than $\omega_{\rho}$ for the P wave double-heavy baryons and thus $\rho$ modes are dominant. This is shown in Fig.\ref{GZAI_MASS_DEP} and Fig.\ref{GZAI_OMEGA_PROB}. One sees that the $(1/2)^2$, $(3/2^-)^2$, $(5/2^-)$ $\lambda$ modes and the $(1/2^-,3/2^-)$ $\rho$ modes split in the heavy quark region in Fig.\ref{GZAI_MASS_DEP}, and the $\rho$ modes become dominant for the lowest states at $m_Q\geq m_c$ in Fig.\ref{GZAI_OMEGA_PROB}. 

\begin{figure*}[htbp]
  \begin{center}
    \begin{tabular}{c}

      \begin{minipage}{1.0\hsize}
        \begin{center}
   \includegraphics[trim = 40 435 10 0,clip, width=17.6cm]{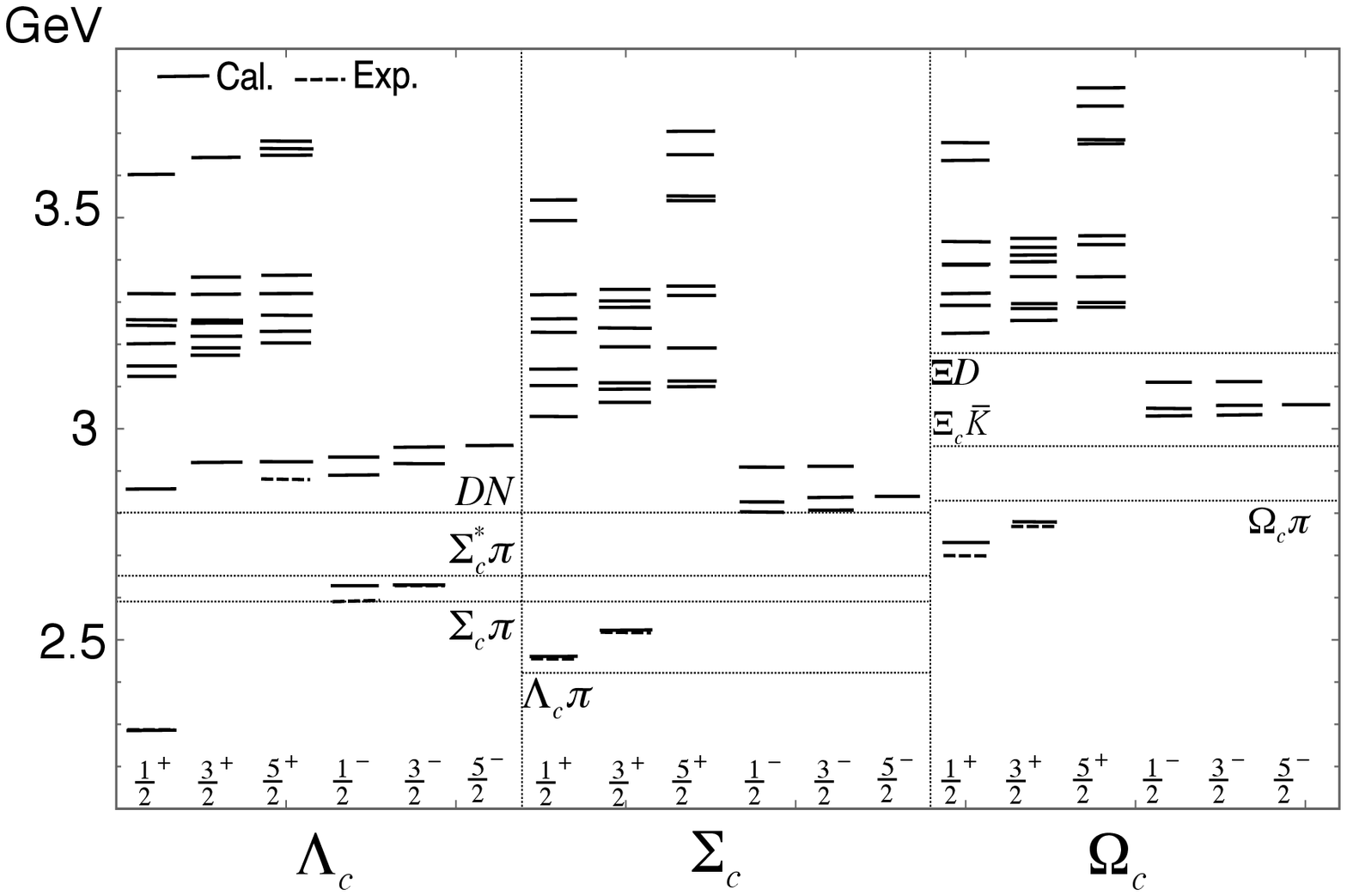}
        \end{center}
      \end{minipage}
    \end{tabular}
    \caption{Calculated energy spectra of $\Lambda_c$, $\Sigma_c$, $\Omega_c$ for $1/2^{+}$, $3/2^{+}$, $5/2^{+}$,  $1/2^{-}$, $3/2^{-}$, $5/2^{-}$(solid line) together with experimental data (dashed line). Several thresholds are also shown by doted line.}
    \label{FGcharm}
  \end{center}
\end{figure*}

\subsection{Heavy baryons in the heavy quark limit}
In this subsection, we investigate the behavior of the single-heavy baryons in the heavy quark limit. We decompose the wave functions of the P-wave single-heavy baryons into the parts with different light spin component $j$ as 
\begin{eqnarray}
      \Phi_{\Lambda_Q}^{J=1/2,M}(\bm{\rho},\bm{\lambda})&=& \phi^{J=1/2,M}_{\Lambda_Q,j=0}(\bm{\rho},\bm{\lambda})\nonumber\\
      &+& \phi^{J=1/2,M}_{\Lambda_Q,j=1}(\bm{\rho},\bm{\lambda}) \label{Lambda_1_2_j_base}
        \end{eqnarray}
        \begin{eqnarray}
      \Phi_{\Lambda_Q}^{J=3/2,M}(\bm{\rho},\bm{\lambda})&=& \phi^{J=3/2,M}_{\Lambda_Q,j=1}(\bm{\rho},\bm{\lambda})\nonumber\\
      &+& \phi^{J=3/2,M}_{\Lambda_Q,j=2}(\bm{\rho},\bm{\lambda}) 
        \end{eqnarray}
      \begin{eqnarray}
      \Phi_{\Sigma_Q}^{J=1/2,M}(\bm{\rho},\bm{\lambda})&=& \phi^{J=1/2,M}_{\Sigma_Q,j=0}(\bm{\rho},\bm{\lambda})\nonumber\\
      &+& \phi^{J=1/2,M}_{\Sigma_Q,j=1}(\bm{\rho},\bm{\lambda}) 
        \end{eqnarray}
      \begin{eqnarray}
      \Phi_{\Sigma_Q}^{J=3/2,M}(\bm{\rho},\bm{\lambda})&=& \phi^{J=3/2,M}_{\Sigma_Q,j=1}(\bm{\rho},\bm{\lambda})\nonumber\\
      &+& \phi^{J=3/2,M}_{\Sigma_Q,j=2}(\bm{\rho},\bm{\lambda}). \label{Sigma_3_2_j_base}
  \end{eqnarray}
Here, we take into account only the channel $c=3$ of the Jacobi coordinates, given in Fig.\ref{jacobi}. The relation between the representation Eq.(\ref{Lambda_1_2_j_base})-(\ref{Sigma_3_2_j_base}) and Eq.(\ref{total-wave-function}) is shown in Appendix \ref{appendix_A}.
The heavy quark mass dependences of the probabilities of each $j$ state are shown in the Figs.\ref{lambda12}-\ref{sigma32} (See Appendix \ref{appendix_A} for the definition). The mixings between $j=0$ and $j=1$ or $j=1$ and $j=2$ above 1 GeV are negligible for the first state of $\Lambda_Q(1/2^-)$ and $\Lambda_Q(3/2^-)$ and the third state of $\Sigma_Q(1/2^-)$ and $\Sigma_Q(3/2^-)$, which correspond to red lines in Figs \ref{lambda12}-\ref{lambda32} and green lines in Figs.\ref{sigma12}-\ref{sigma32}. This is because these states  are isolated from the other states, as is shown in Fig.\ref{LAMBDA_SIGMA_MASS_DEP}. For the other state, two different $j$ components (five $\lambda$-modes of $\Sigma_Q$ and five $\rho$-modes of $\Lambda_Q$) still mix in the charm and bottom mass region, because they lie close to each other within 50 MeV (See Fig.\ref{LAMBDA_SIGMA_MASS_DEP}). Above $m_Q=$14 GeV, one sees no mixing between different $j$ components.  In summary, one finds that the second $1/2^-$ state of $\Lambda_Q$ and the first $1/2^-$ state of 1/2$^-$ of $\Sigma_Q$ are the $j$=0 singlet state. All the other belong to doublets, $(1/2_1^-, 3/2_1^-)$, $(1/2_3^-, 3/2_2^-)$ and  $(3/2_3^-, 5/2_1^-)$ for $\Lambda_Q$ and $(1/2_2^-, 3/2_1^-)$, $(3/2_2^-, 5/2_1^-)$ and $(1/2_3^-, 3/2_2^-)$ for $\Sigma_Q$, as is shown in Fig.\ref{LAMBDA_SIGMA_MASS_DEP}. 

We next discuss the positive parity states. We focus on the first six positive parity state of single-heavy baryons, corresponding to the states below 3.0 GeV in the charm sector (See Fig.\ref{FGcharm}).They consist of the S-wave (($L$,$\ell$)=(0,0)) component, the (1,1) component, the (2,0) component ($\rho$-mode) and the (0,2) component ($\lambda$-mode). Figs.\ref{lambda12K1}-\ref{sigma32K1} show the probabilities of the each component in the total wave function. One sees that one component becomes dominant avove $m_Q=$1 GeV. The (0,0) component is dominant for $\Lambda_Q(1/2_1^+)$, $\Lambda_Q(1/2_2^+)$, $\Sigma_Q(1/2_1^+)$, $\Sigma_Q(1/2_2^+)$ and (2,0)  component ($\lambda$-mode) is dominant for $\Lambda_Q(3/2_1^+)$, $\Lambda_Q(5/2_1^+)$ above 1 GeV (See Figs.\ref{lambda12K1}-\ref{sigma32K1}). The lowest six states in the heavy quark region can be written as follows.
\begin{eqnarray}
 \Phi_{\Lambda_Q}^{J_n=1/2_1,M}(\bm{\rho},\bm{\lambda})&=&  \phi^{J_n=1/2_1^+,M}_{\Lambda_Q,j=0}(\bm{\rho},\bm{\lambda})  \label{HQS_positive1} 
   \end{eqnarray}
   \begin{eqnarray}
 \Phi_{\Lambda_Q}^{J_n=1/2_2,M}(\bm{\rho},\bm{\lambda})&=&  \phi^{J_n=1/2_2,M}_{\Lambda_Q,j=0}(\bm{\rho},\bm{\lambda}) 
   \end{eqnarray}
   \begin{eqnarray}
 \Phi_{\Lambda_Q}^{J_n=3/2_1,M}(\bm{\rho},\bm{\lambda})&=&  \phi^{J_n=3/2_1,M}_{\Lambda_Q,j=2}(\bm{\rho},\bm{\lambda}) 
   \end{eqnarray}
   \begin{eqnarray}
 \Phi_{\Lambda_Q}^{J_n=5/2_1,M}(\bm{\rho},\bm{\lambda}) &=&  \phi^{J_n=5/2_1,M}_{\Lambda_Q,j=2}(\bm{\rho},\bm{\lambda})
   \end{eqnarray}
   \begin{eqnarray}
 \Phi_{\Sigma_Q}^{J_n=1/2_1,M}(\bm{\rho},\bm{\lambda})   &=& \phi^{J_n=1/2_1,M}_{\Sigma_Q,j=1}(\bm{\rho},\bm{\lambda})
   \end{eqnarray}
   \begin{eqnarray}
 \Phi_{\Sigma_Q}^{J_n=3/2_1,M}(\bm{\rho},\bm{\lambda})   &=& \phi^{J_n=3/2_1,M}_{\Sigma_Q,j=1}(\bm{\rho},\bm{\lambda})
 \label{HQS_positive6}
\end{eqnarray}
where we use Eq.(\ref{HQS_Base}) to transform the bases. There are two doublet pairs ($\Lambda_Q(3/2_1^+)$, $\Lambda_Q(5/2_1^+)$) ($j=2$), ($\Sigma_Q(1/2_1^+)$, $\Sigma_Q(3/2_1^+)$) ($j=1$) and two singlet states $\Lambda_Q(1/2_1^+)$, $\Lambda_Q(1/2_2^+)$ in the heavy quark limit. Mixings of different $j$ components of the wave function are negligible even in the charm quark region.

\section{Summary}
We have studied the spectrum of the single- and double-heavy baryons and discussed their structures within the framework of a constituent quark model. The potential parameters are determined so as to reproduce the energies of the lowest states $\Lambda$($1/2^+$), $\Sigma$($1/2^+$), $\Sigma$($3/2^+$), $\Lambda$($1/2^-$), $\Lambda$($3/2^-$), $\Lambda_c$($1/2^+$) and $\Lambda_b$($1/2^+$). In the analysis of the baryon wave functions, we have focused on the two characteristic excited modes and investigated the their probabilities as functions of the heavy quark mass. To obtain the precise energy eigenvalues of excited states, we employ the gaussian expansion method, which is one of the best method for three and four body bound states. We have obtained the followings:\\

(1) Masses of the known $\Lambda_Q$, $\Sigma_Q$ and $\Omega_Q$ are in good agreement with the observed data within 50 MeV. Then, we predicted that observed $\Sigma_c$(2800) can be assigned to $1/2_1^-$, $3/2_1^-$, $1/2_2^-$, $3/2_2^-$ and $5/2_1^-$ state, and $\Lambda_c$ to $3/2_1^+$, $5/2_1^-$, $1/2_2^-$, $1/2_3^-$ and $3/2_2^-$ and $3/2_3^-$.\\

(2) In the heavy quark limit, we find six doublets and two singlets for the P-wave single-heavy baryons (See Fig.\ref{LAMBDA_SIGMA_MASS_DEP}) and two doublets and two singlets for the first six states of positive parity single-heavy baryons. In the charm sector, the mass differences of these heavy quark spin-doublets are less than 30 [MeV] and in the bottom sector, the differences reduce to less than 10 [MeV].\\

(3) For the double-heavy baryons, we predict that the mass of the ground $\Xi_{cc}$ state is $\Xi_{cc}(3685)$. This result is consistent with the recent Lattice QCD calculations within 50 MeV. Experimentally, it was reported that a double- charmed baryon was found at the mass 3512 MeV \cite{Moinester:2002uw}. But other experimental groups, LHC and  Belle, have not yet succeeded in the observing the state.\\

(4) We have investigated the dependences on the heavy quark mass $m_Q$ of the $\lambda$ and $\rho$ modes to see the features of the negative parity states. Mixings of the $\rho$ and $\lambda$ modes are suppressed and only one mode dominates. This is because the spin-spin interaction which mainly causes the mixing becomes small in the heavy quark region. It is a future problem to clarify what physical quantities sensitive the differences of the two modes. One possibility is decay patterns. It is conjectured that the $\lambda$-mode states decay dominantly to a light baryon and a heavy meson, while the $\rho$-mode states decay mostly into a light meson and a heavy baryon. Further studies of the decays and productions of these heavy baryons will be useful to verify more on these structures.  

\section*{Acknowledgments}
The authors would like to thank Drs. Shigehiro Yasui, Hiroyuki Noumi for valuable discussions. 
This work was supported in part by JSPS KAKENHI numbers, 25247036, 24250294, 26400273. 
T.Y. acknowledges the Junior Research Associate scholarship at RIKEN. 
The numerical calculations were carried out on SR16000 at YITP in Kyoto University.

\appendix
\def\thesection{\Alph{section}}
\section{ The transformation of the bases} \label{appendix_A}
We discuss the wave function in the heavy quark limit in this appendix. For the single-heavy baryons, we take only the channel c=3  of the Jscobi coordinate given in Fig.\ref{jacobi}. The P-wave wave functions of the $\Lambda_Q$ and $\Sigma_Q$ baryons are given by the sum of the $\lambda$-mode ($^{2S+1}{\lambda}=$$^2\lambda$, $^4\lambda$) and $\rho$-mode ($^{2S+1}\rho=$$^2\rho$, $^4\rho$) components as follows.
\begin{eqnarray}
      \Phi_{\Lambda_Q}^{JM}(\bm{\rho},\bm{\lambda})&=&\psi^{\Lambda_Q}_{^2\rho}\sum_{(n,N)}C^{^2\rho}_{n,N}\phi_{n,N}(\rho,\lambda) \nonumber \\
      &+&\psi^{\Lambda_Q}_{^4\rho}\sum_{(n,N)}C^{^4\rho}_{n,N}\phi_{n,N}(\rho,\lambda)  \nonumber \\
      &+&\psi^{\Lambda_Q}_{^2\lambda}\sum_{(n,N)}C^{^2\lambda}_{n,N}\phi_n(\rho,\lambda) \label{LAMBDA_WAVE}
  \end{eqnarray}
\begin{eqnarray}
      \Phi_{\Sigma_Q}^{JM}(\bm{\rho},\bm{\lambda})&=&\psi^{\Sigma_Q}_{^2\lambda}\sum_{(n,N)}C^{^2\lambda}_{n,N}\phi_n(\rho,\lambda) \nonumber \\
      &+&\psi^{\Sigma_Q}_{^4\lambda}\sum_{(n,N)}C^{^4\lambda}_{n}\phi_{n,N}(\rho,\lambda)  \nonumber \\
      &+&\psi^{\Sigma_Q}_{^2\rho}\sum_{(n,N)}C^{^2\rho}_{n,N}\phi_{n,N}(\rho,\lambda). \label{SIGMA_WAVE}
  \end{eqnarray}
  Here we extract the parts of the spin and orbital angular momenta for each mode as
\begin{eqnarray}
      \psi^{\Lambda_Q}_{^2\rho}&=&[X_{S=1/2,1}[Y_{\ell=1}(\bm{\hat{\rho}})Y_{L=0}(\bm{\hat{\lambda}})]_{I=1}]_{JM}  \label{trans_1}  \\
      \psi^{\Lambda_Q}_{^4\rho}&=&[X_{S=3/2,1}[Y_{\ell=1}(\bm{\hat{\rho}})Y_{L=0}(\bm{\hat{\lambda}})]_{I=1}]_{JM}  \\
      \psi^{\Lambda_Q}_{^2\lambda}&=&[X_{S=1/2,0}[Y_{\ell=0}(\bm{\hat{\rho}})Y_{L=1}(\bm{\hat{\lambda}})]_{I=1}]_{JM} \\
       \psi^{\Sigma_Q}_{^2\lambda}&=&[X_{S=1/2,1}[Y_{\ell=0}(\bm{\hat{\rho}})Y_{L=1}(\bm{\hat{\lambda}})]_{I=1}]_{JM} \\
      \psi^{\Sigma_Q}_{^4\lambda}&=&[X_{S=3/2,1}[Y_{\ell=0}(\bm{\hat{\rho}})Y_{L=1}(\bm{\hat{\lambda}})]_{I=1}]_{JM}  \\
      \psi^{\Sigma_Q}_{^2\rho}&=&[X_{S=1/2,0}[Y_{\ell=1}(\bm{\hat{\rho}})Y_{L=0}(\bm{\hat{\lambda}})]_{I=1}]_{JM}
      \label{trans_2} 
\end{eqnarray}
Then the corresponding radial parts are expanded by the Gaussian basis as
 \begin{equation}
\phi_{\it{n},N}(\rho,\lambda)= N_{n\ell}N_{NL}\\\rho^{\ell}e^{-\beta_{n}\rho^2}\lambda^Le^{{-\gamma_N
\lambda^2}},
\end{equation}
where $N_{nl}$($N_{NL}$) is the normalization constant. As is discussed in Sec.III.C, the light spin component $j$ is conserved in the heavy quark limit. Therefore, we transform the bases into those which diagonalize $j$. We use Eq.(\ref{HQS_Base}) to transform the bases, and obtain
\begin{eqnarray}
      \psi^{\Lambda_Q}_{^2\rho}&=&\sqrt{\frac{1}{3}}\psi_{j=0,s=1}-\sqrt{\frac{2}{3}}\psi_{j=1,s=1} \\
      \psi^{\Lambda_Q}_{^4\rho}&=&\sqrt{\frac{2}{3}}\psi_{j=0,s=1}+\sqrt{\frac{1}{3}}\psi_{j=1,s=1} \\
      \psi^{\Lambda_Q}_{^2\lambda}&=&-\psi_{j=1,s=0}\\
      \psi^{\Sigma_Q}_{^2\lambda}&=&\sqrt{\frac{1}{3}}\psi_{j=0,s=1}-\sqrt{\frac{2}{3}}\psi_{j=1,s=1} \\
      \psi^{\Sigma_Q}_{^4\lambda}&=&\sqrt{\frac{2}{3}}\psi_{j=0,s=1}+\sqrt{\frac{1}{3}}\psi_{j=1,s=1}\\
       \psi^{\Sigma_Q}_{^2\rho}&=&-\psi_{j=1,s=0}.
\end{eqnarray}
for $J=1/2^-$ and
\begin{eqnarray}
      \psi^{\Lambda_Q}_{^2\rho}&=&-\sqrt{\frac{1}{6}}\psi_{j=1,s=1}+\sqrt{\frac{5}{6}}\psi_{j=2,s=1} \\
      \psi^{\Lambda_Q}_{^4\rho}&=&-\sqrt{\frac{5}{6}}\psi_{j=1,s=1}-\sqrt{\frac{1}{6}}\psi_{j=2,s=1} \\
      \psi^{\Lambda_Q}_{^2\lambda}&=&\psi_{j=1,s=0}\\
      \psi^{\Sigma_Q}_{^2\lambda}&=&-\sqrt{\frac{1}{6}}\psi_{j=1,s=1}+\sqrt{\frac{5}{6}}\psi_{j=2,s=1} \\
      \psi^{\Sigma_Q}_{^4\lambda}&=&-\sqrt{\frac{5}{6}}\psi_{j=1,s=1}-\sqrt{\frac{1}{6}}\psi_{j=2,s=1}\\
       \psi^{\Sigma_Q}_{^2\rho}&=&\psi_{j=1,s=0}.
\end{eqnarray}
for $J=3/2^-$, where 
\begin{eqnarray}
\scalebox{0.9}{$\displaystyle\psi_{j,s}= [[[\chi_{1/2}(q)\chi_{1/2}(q)]_{s}[Y(\hat{\bm{\rho}})_{\ell}Y(\hat{\bm{\lambda}})_{L}]_I]_j\chi_{1/2}(Q)]_{J}$}
\end{eqnarray}
By using Eq.(\ref{trans_1})-(\ref{trans_2}),  Eq.(\ref{LAMBDA_WAVE}) and Eq(\ref{SIGMA_WAVE})  is transformed into the bases which is characterized by $j$.
\begin{itemize}
\item $\Lambda_Q(1/2^-,3/2^-)$
\begin{eqnarray}
 \phi^{J=1/2,M}_{\Lambda_Q,j=0}(\bm{\rho},\bm{\lambda})&=&\psi_{j=0,s=1}\left(\sqrt{\frac{1}{3}}\sum_{(n,N)}C^{^2\rho}_{n,N}\phi_{n,N}(\rho,\lambda)\right.\nonumber\\
&+&\sqrt{\frac{2}{3}}\sum_{(n,N)}C^{^4\rho}_{n,N}\phi_{n,N}(\rho,\lambda)\Biggr)\nonumber\\
&-&\psi_{j=0,s=0}\sum_{(n,N)}C^{^2\lambda}_{n,N}\phi_{n,N}(\rho,\lambda)
  \end{eqnarray}
  \begin{eqnarray}
        \phi^{J=1/2,M}_{\Lambda_Q,j=1}(\bm{\rho},\bm{\lambda})&=&\psi_{j=1,s=1}\left(-\sqrt{\frac{2}{3}}\sum_{(n,N)}C^{^2\rho}_{n,N}\phi_{n,N}(\rho,\lambda)\right.\nonumber\\
 &+&\sqrt{\frac{1}{3}}\sum_{(n,N)}C^{^4\rho}_{n,N}\phi_{n,N}(\rho,\lambda)\Biggr)
  \end{eqnarray}
  \begin{eqnarray}
      \phi^{J=3/2,M}_{\Lambda_Q,j=1}(\bm{\rho},\bm{\lambda})&=&\psi_{j=1,s=1}\left(-\sqrt{\frac{1}{6}}\sum_{(n,N)}C^{^2\rho}_{n,N}\phi_{n,N}(\rho,\lambda)\right.\nonumber\\
&-&\sqrt{\frac{5}{6}}\sum_{(n,N)}C^{^4\rho}_{n,N}\phi_{n,N}(\rho,\lambda)\Biggr)\nonumber\\
&+&\psi_{j=1,s=0}\sum_{(n,N)}C^{^2\lambda}_{n,N}\phi_{n,N}(\rho,\lambda) 
 \end{eqnarray}
  \begin{eqnarray}
        \phi^{J=3/2,M}_{\Lambda_Q,j=2}(\bm{\rho},\bm{\lambda})&=&\psi_{j=2,s=1}\left(\sqrt{\frac{5}{6}}\sum_{(n,N)}C^{^2\rho}_{n,N}\phi_{n,N}(\rho,\lambda)\right. \nonumber \\
 &-&\sqrt{\frac{1}{6}}\sum_{(n,N)}C^{^4\rho}_{n,N}\phi_{n,N}(\rho,\lambda)\Biggr)
  \end{eqnarray}
  
\item $\Sigma_Q(1/2^-,3/2^-)$
\begin{eqnarray}
      \phi^{J=1/2,M}_{\Sigma_Q,j=0}(\bm{\rho},\bm{\lambda})&=&\psi_{j=0,s=1}\left(\sqrt{\frac{1}{3}}\sum_{(n,N)}C^{^2\lambda}_{n,N}\phi_{n,N}(\rho,\lambda)\right. \nonumber \\
 &+&\sqrt{\frac{2}{3}}\sum_{(n,N)}C^{^4\lambda}_{n,N}\phi_{n,N}(\rho,\lambda)\Biggr)\nonumber \\ 
 &-&\psi_{j=0,s=0}\sum_{(n,N)}C^{^2\rho}_{n,N}\phi_{n,N}(\rho,\lambda)
  \end{eqnarray}
  \begin{eqnarray}
        \phi^{J=1/2,M}_{\Sigma_Q,j=1}(\bm{\rho},\bm{\lambda})&=&\psi_{j=1,s=1}\left(-\sqrt{\frac{2}{3}}\sum_{(n,N)}C^{^2\lambda}_{n,N}\phi_{n,N}(\rho,\lambda)\right.\nonumber\\
        &&\left.+\sqrt{\frac{1}{3}}\sum_{(n,N)}C^{^4\lambda}_{n,N}\phi_{n,N}(\rho,\lambda)\right)
  \end{eqnarray}
\begin{eqnarray}
      \phi^{J=3/2,M}_{\Sigma_Q,j=1}(\bm{\rho},\bm{\lambda})&=&\psi_{j=1,s=1}\left(-\sqrt{\frac{1}{6}}\sum_{(n,N)}C^{^2\lambda}_{n,N}\phi_{n,N}(\rho,\lambda)\right.\nonumber\\
      &-&\sqrt{\frac{5}{6}}\sum_{(n,N)}C^{^4\lambda}_{n,N}\phi_{n,N}(\rho,\lambda)\Biggr)\nonumber\\
      &+&\psi_{j=1,s=0}\sum_{(n,N)}C^{^2\rho}_{n,N}\phi_{n,N}(\rho,\lambda)
       \end{eqnarray}
  \begin{eqnarray}
        \phi^{J=3/2,M}_{\Sigma_Q,j=2}(\bm{\rho},\bm{\lambda})&=&\psi_{j=2,s=1}\left(\sqrt{\frac{5}{6}}\sum_{(n,N)}C^{^2\lambda}_{n,N}\phi_{n,N}(\rho,\lambda)\right.\nonumber\\
      &&\left.  -\sqrt{\frac{1}{6}}\sum_{(n,N)}C^{^4\lambda}_{n,N}\phi_{n,N}(\rho,\lambda)\right)
  \end{eqnarray}
\end {itemize}

\begin{figure*}[htbp]
  \begin{center}
    \begin{tabular}{c}

      \begin{minipage}{0.97\hsize}
        \begin{center}
          \includegraphics[trim = 0 350 0 0,clip, width=12.8cm]{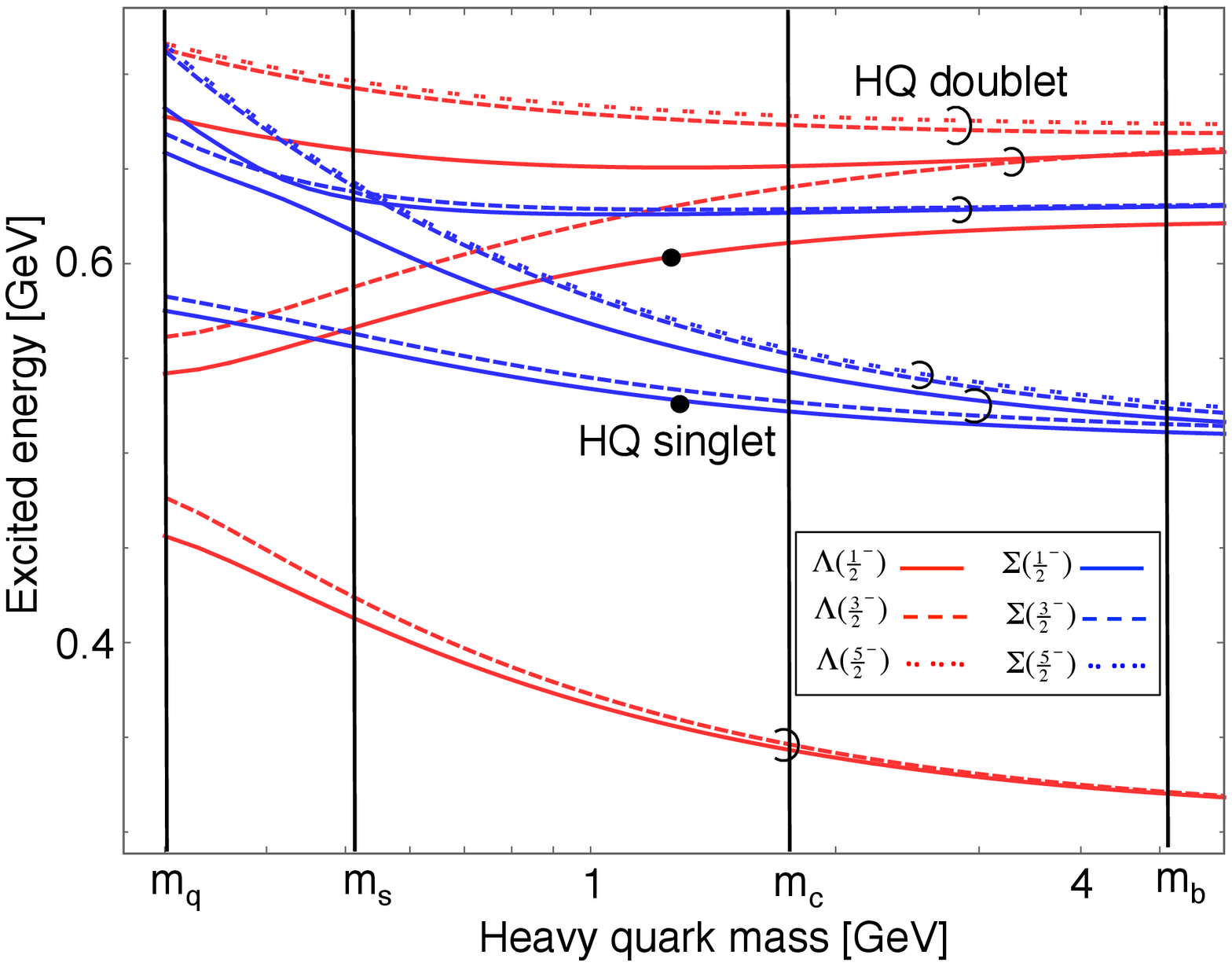}
        \end{center}
      \end{minipage}

    \end{tabular}
    \caption{Heavy quark mass dependence of excited energy of first state, second state and third state for $1/2^{-}$(solid line), $3/2^{-}$(dashed line), $5/2^{-}$(twined line) of $\Lambda_Q$ (red line) and  $\Sigma_Q$ (blue line). Bullet denote heavy quark singlet. The pair within a half circle denote heavy quark doublet.}
    \label{LAMBDA_SIGMA_MASS_DEP}
  \end{center}
\end{figure*}

\begin{figure*}[htbp]
  \begin{center}
    \begin{tabular}{c}

      \begin{minipage}{0.97\hsize}
        \begin{center}
          \includegraphics[trim = 0 370 0 0,clip, width=12.8cm]{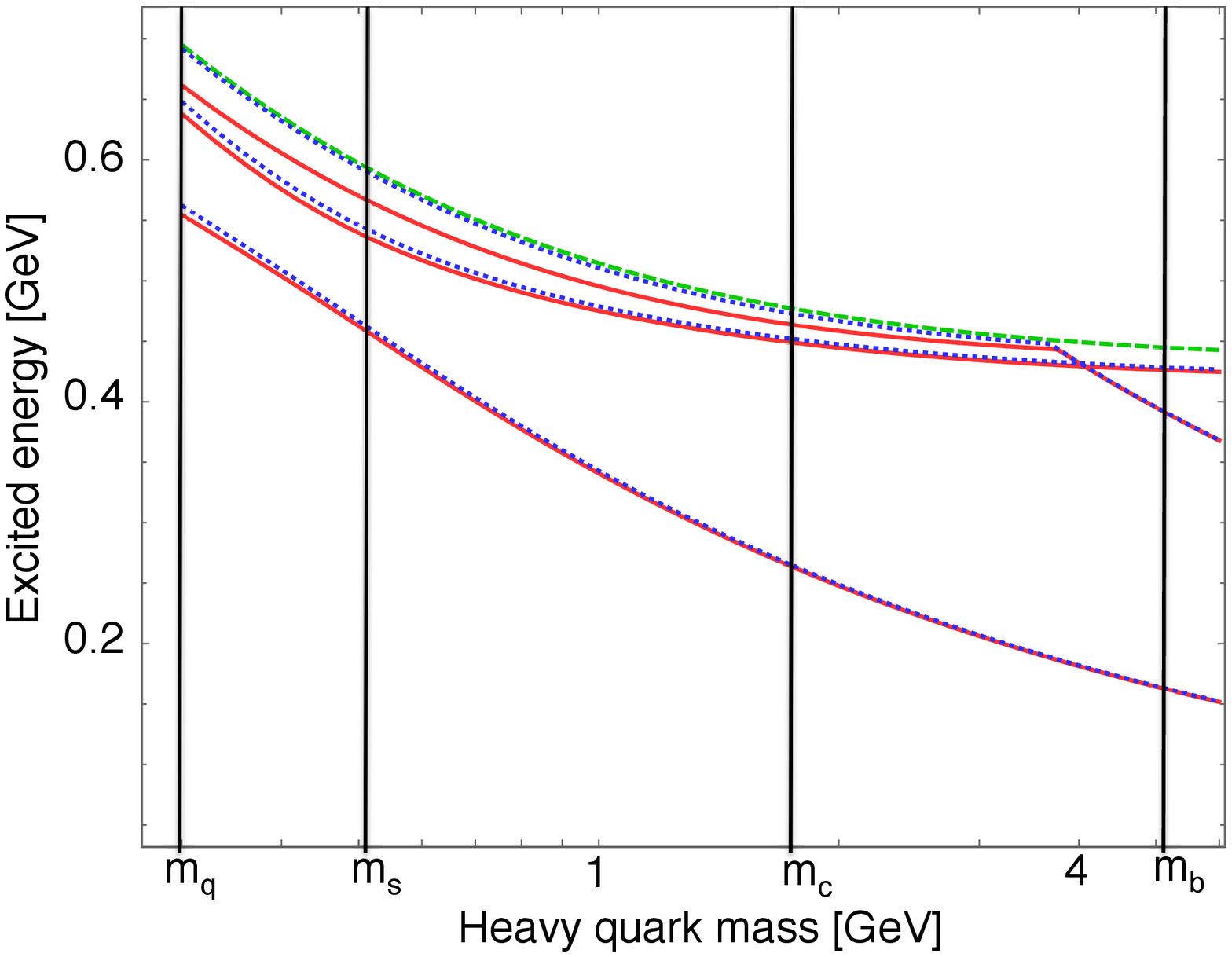}
        \end{center}
      \end{minipage}

    \end{tabular}
    \caption{Heavy quark mass dependence of excited energy of first state, second state and third state for $1/2^{-}$(red solid line), $3/2^{-}$(blue dotted line), $5/2^{-}$(green dashed line) of $\Xi_{QQ}$. }
    \label{GZAI_MASS_DEP}
  \end{center}
\end{figure*}

\begin{figure*}[htbp]
  \begin{center}
    \begin{tabular}{c}

      \begin{minipage}{0.95\hsize}
       \begin{center}
          \includegraphics[trim = 10 390 0 0 ,clip, width=12.0cm]{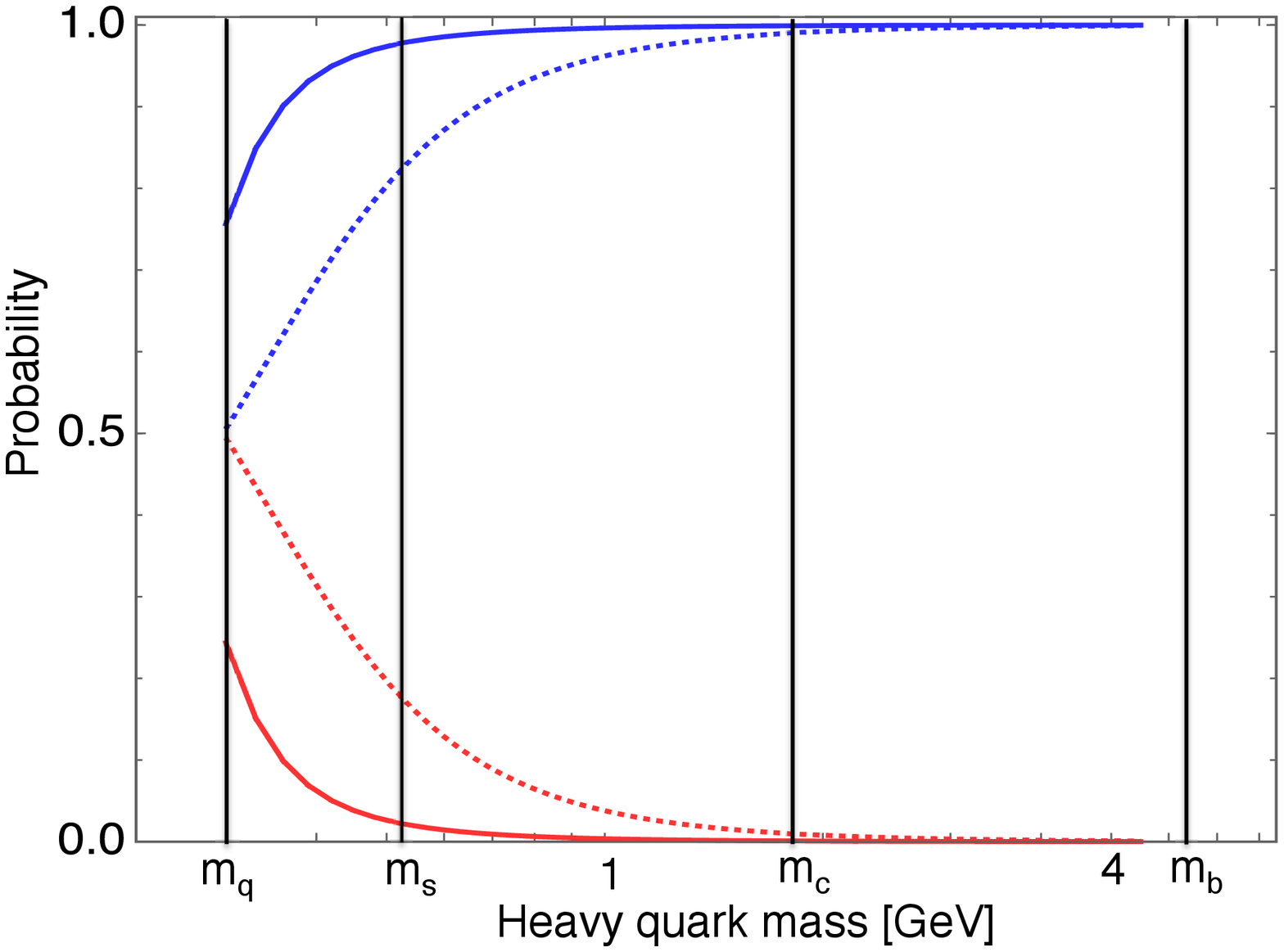}
        \end{center}
      \end{minipage}
      
\end{tabular}
    \caption{The prbability of $\lambda$ mode (blue line) and $\rho$ mode (red line)  of $\frac{1}{2}^-$ for $\Sigma_Q$ (dotted line), $\Lambda_Q$ (Solid line).}
    \label{LAMBDA_SIGMA_PROB}
  \end{center}
\end{figure*}

\begin{figure*}[htbp]
  \begin{center}
    \begin{tabular}{c}

      \begin{minipage}{0.95\hsize}
        \begin{center}
          \includegraphics[trim = 10 390 0 0 ,clip, width=12.0cm]{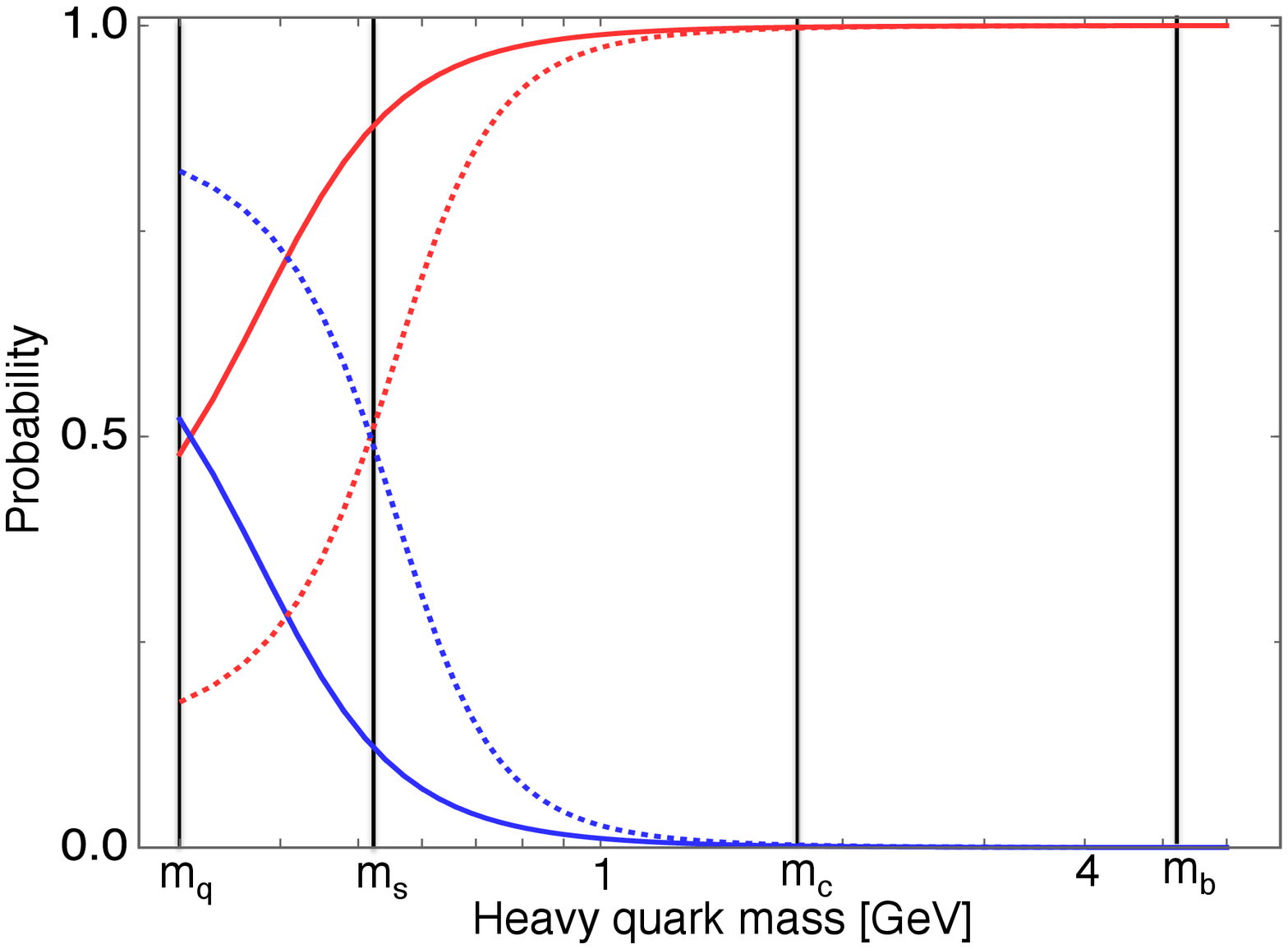}
        \end{center}
     \end{minipage}
      
\end{tabular}
    \caption{The prbability of $\lambda$ mode (blue line) and $\rho$ mode (red line) of $\frac{1}{2}^-$ for $\Xi_{QQ}$ (Solid line) and $\Omega_{QQ}$ (dotted line).}
    \label{GZAI_OMEGA_PROB}
  \end{center}
\end{figure*}

\begin{figure*}[htbp]
  \begin{center}
    \begin{tabular}{c}

      \begin{minipage}{0.97\hsize}
        \begin{center}
          \includegraphics[trim = 0 360 0 50,clip, width=12.8cm]{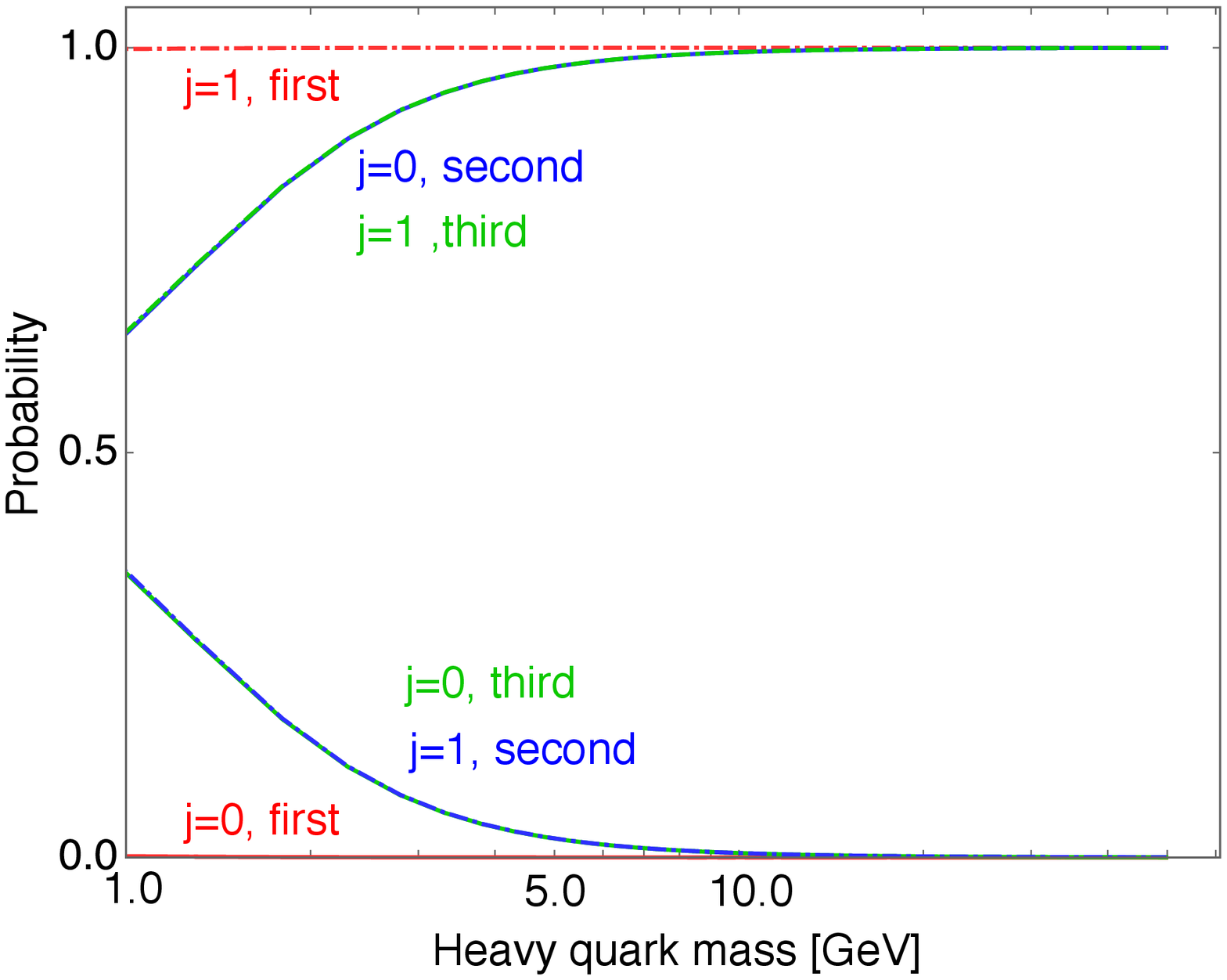}
        \end{center}
      \end{minipage}

    \end{tabular}
    \caption{The probabilities of $j$=0 (Solid line) and $j$=1 (Chain line) for $\Lambda(1/2^-)$. Red, blue, green lines show the first state, second state and third state respectively.}
    \label{lambda12}
  \end{center}
\end{figure*}

\begin{figure*}[htbp]
  \begin{center}
    \begin{tabular}{c}

      \begin{minipage}{0.97\hsize}
        \begin{center}
          \includegraphics[trim = 0 360 0 50,clip, width=12.8cm]{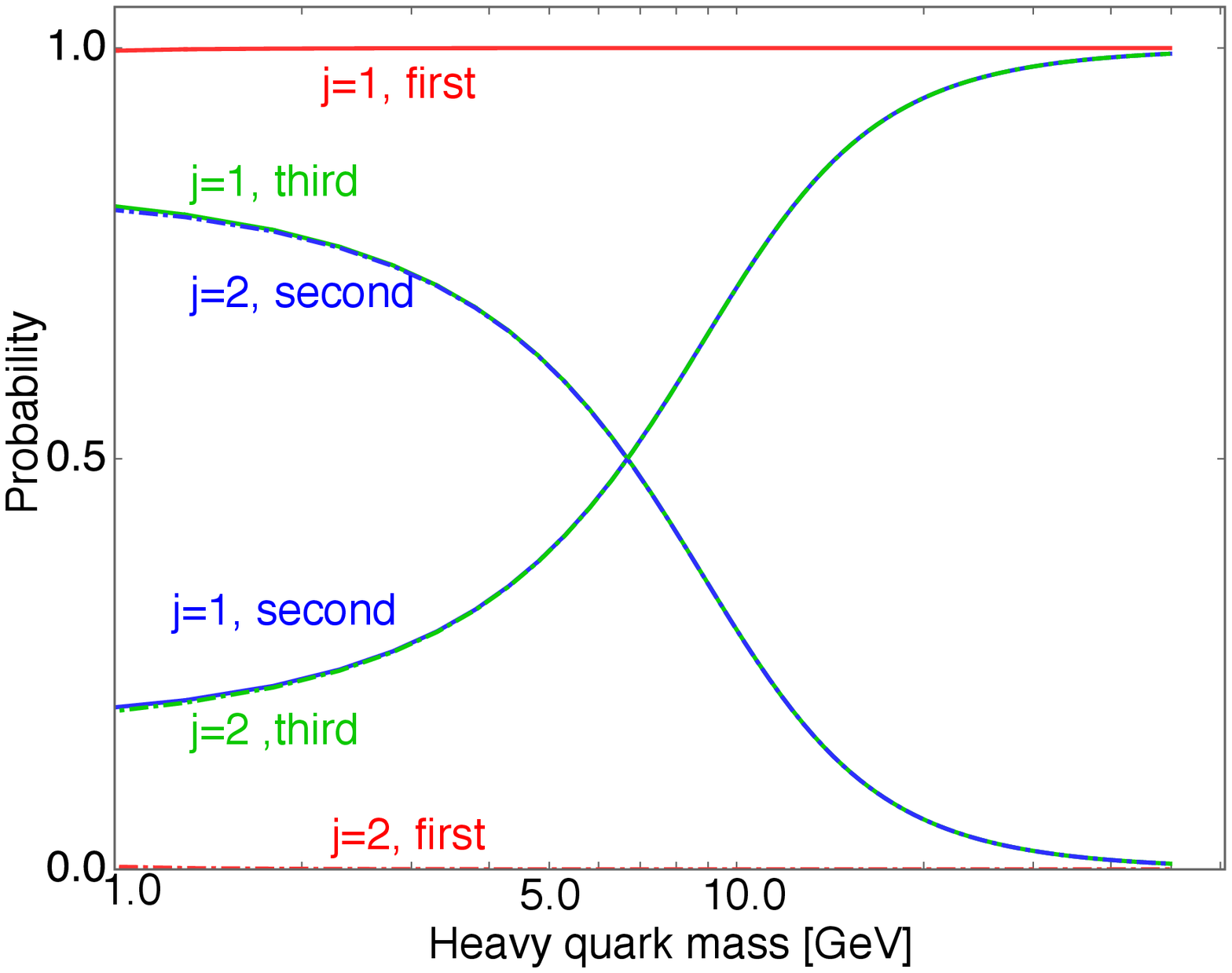}
        \end{center}
      \end{minipage}

    \end{tabular}
    \caption{The probabilities of $j$=1 (Solid line) and $j$=2 (Chain line) for $\Lambda(3/2^-)$. Red, blue, green lines show the first state, second state and third state respectively.}
      \label{lambda32}
  \end{center}
\end{figure*}

\begin{figure*}[htbp]
  \begin{center}
    \begin{tabular}{c}

      \begin{minipage}{0.97\hsize}
        \begin{center}
          \includegraphics[trim = 0 360 0 50,clip, width=12.8cm]{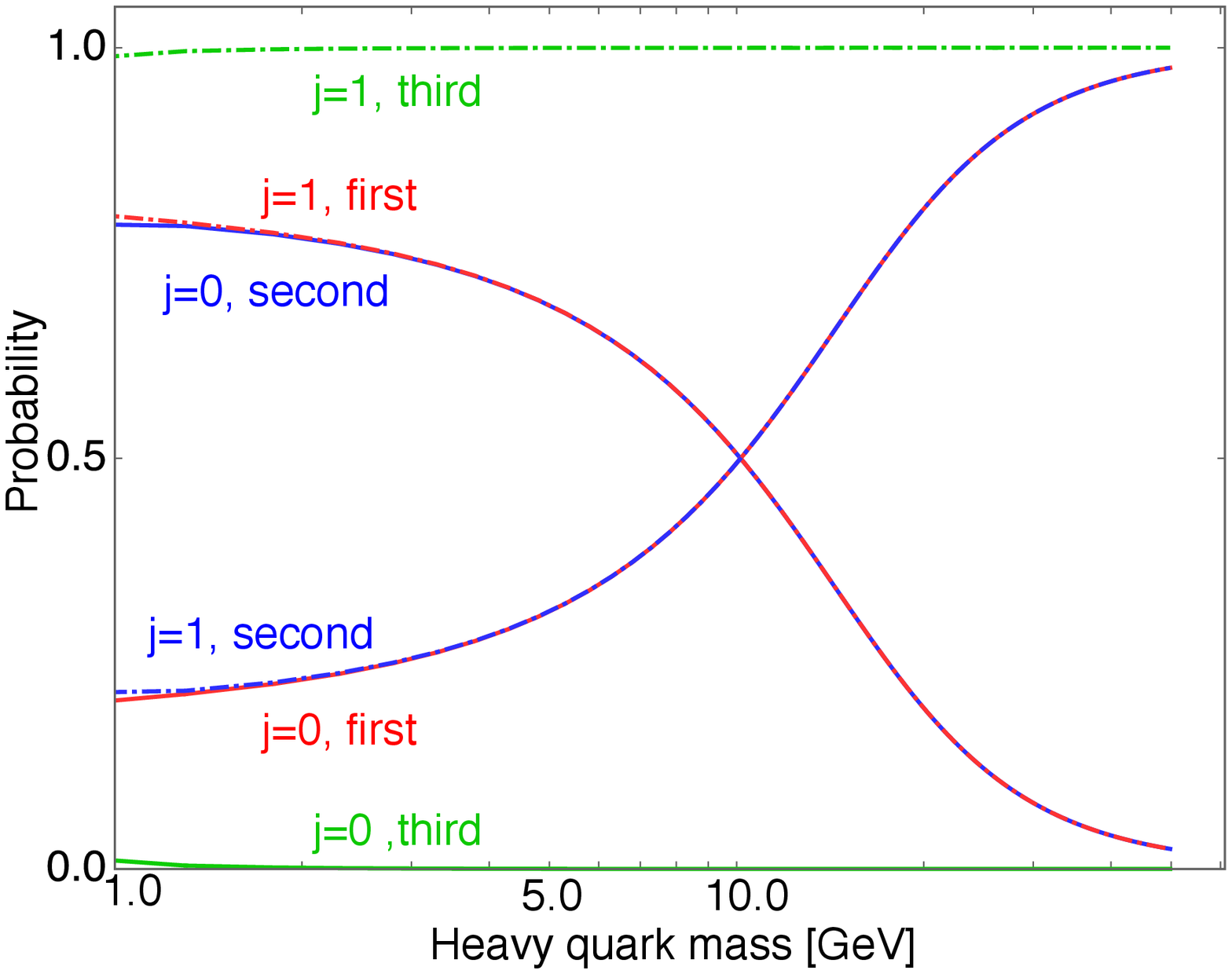}
        \end{center}
      \end{minipage}

    \end{tabular}
    \caption{The probabilities of $j$=0 (Solid line) and $j$=1 (Chain line) for $\Sigma(1/2^-)$. Red, blue, green lines show the first state, second state and third state respectively.}
  \label{sigma12}
  \end{center}
\end{figure*}

\begin{figure*}[htbp]
  \begin{center}
    \begin{tabular}{c}

      \begin{minipage}{0.97\hsize}
        \begin{center}
          \includegraphics[trim = 0 360 0 50,clip, width=12.8cm]{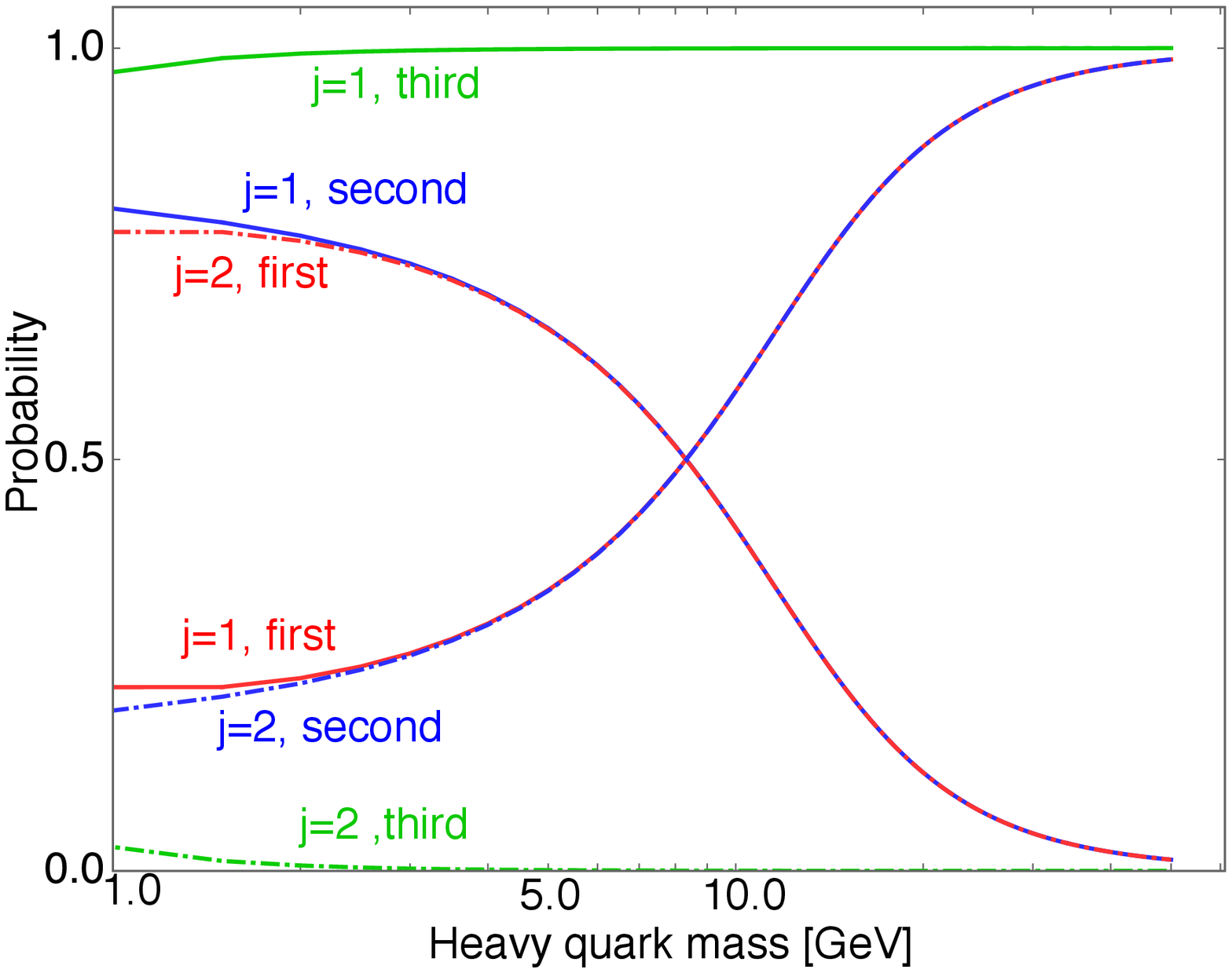}
        \end{center}
      \end{minipage}

    \end{tabular}
    \caption{The probabilities of $j$=1 (Solid line) and $j$=2 (Chain line) for $\Sigma(3/2^-)$. Red, blue, green lines show the first state, second state and third state respectively.}
      \label{sigma32}
  \end{center}
\end{figure*}

\begin{figure*}[htbp]
  \begin{center}
    \begin{tabular}{c}

      \begin{minipage}{0.97\hsize}
        \begin{center}
          \includegraphics[trim = 0 370 0 70,clip, width=12.8cm]{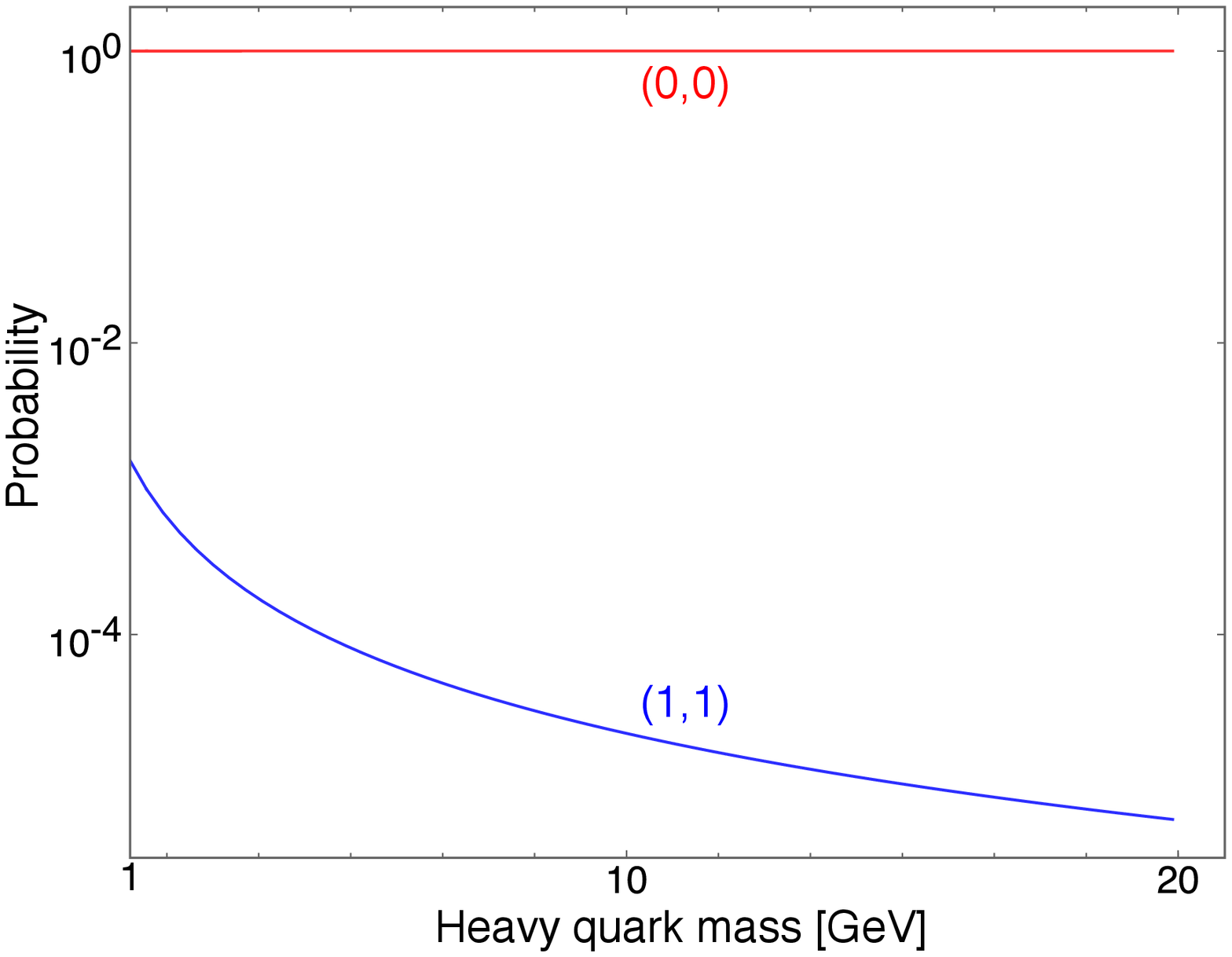}
        \end{center}
      \end{minipage}

    \end{tabular}
    \caption{The heavy quark mass dependence of the probabilities of the S-wave (0.0) component (Red line) and (1,1) component (blue line) for $\Lambda(1/2_1^+)$. }
    \label{lambda12K1}
  \end{center}
\end{figure*}

\begin{figure*}[htbp]
  \begin{center}
    \begin{tabular}{c}

      \begin{minipage}{0.97\hsize}
        \begin{center}
          \includegraphics[trim = 0 370 0 70,clip, width=12.8cm]{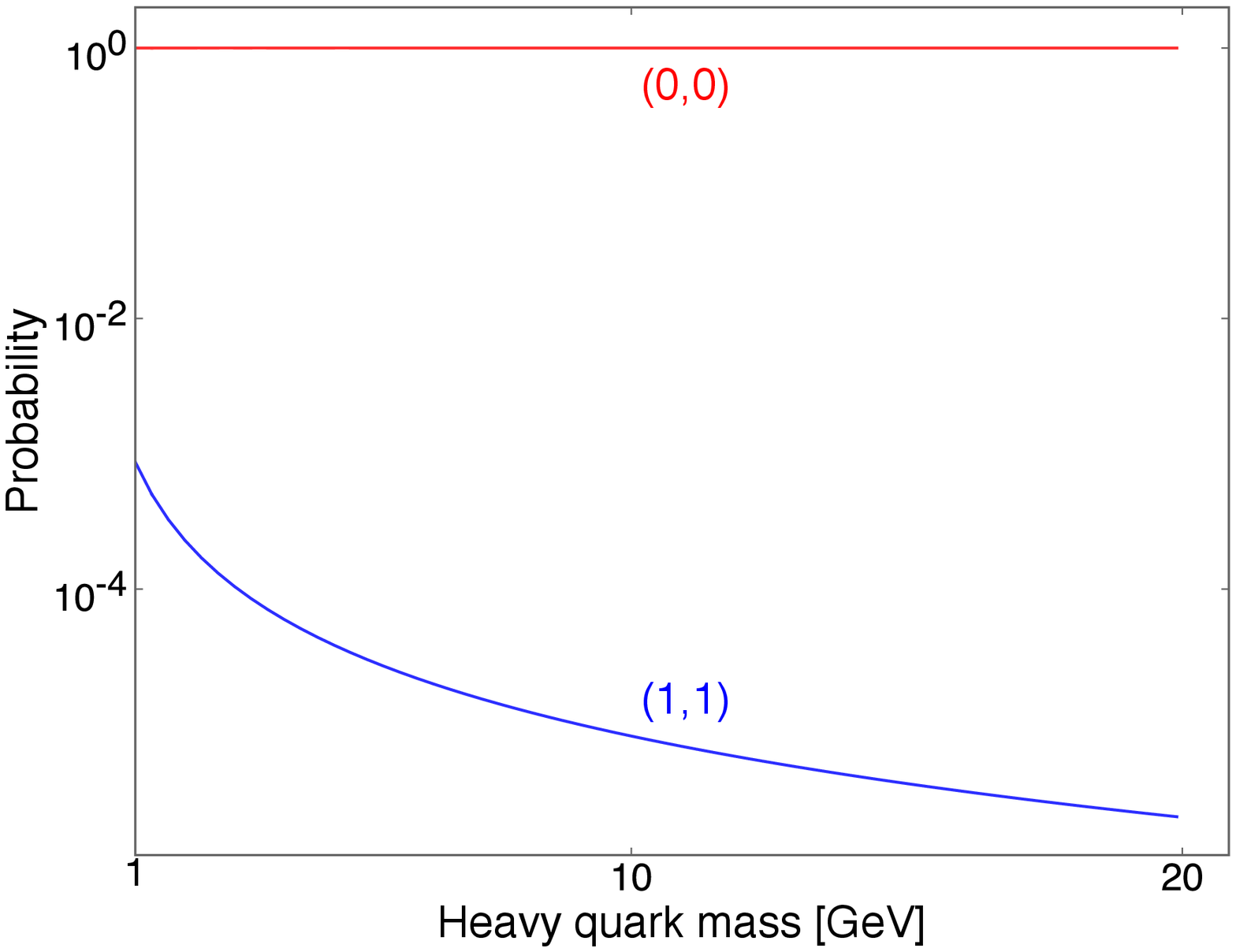}
        \end{center}
      \end{minipage}

    \end{tabular}
    \caption{The heavy quark mass dependences of the probabilities of the S-wave (0.0) component (Red line) and (1,1) component (blue line) for $\Lambda(1/2_2^+)$.}
      \label{lambda12K2}
  \end{center}
\end{figure*}

\begin{figure*}[htbp]
  \begin{center}
    \begin{tabular}{c}

      \begin{minipage}{0.97\hsize}
        \begin{center}
          \includegraphics[trim = 0 370 0 70,clip, width=12.8cm]{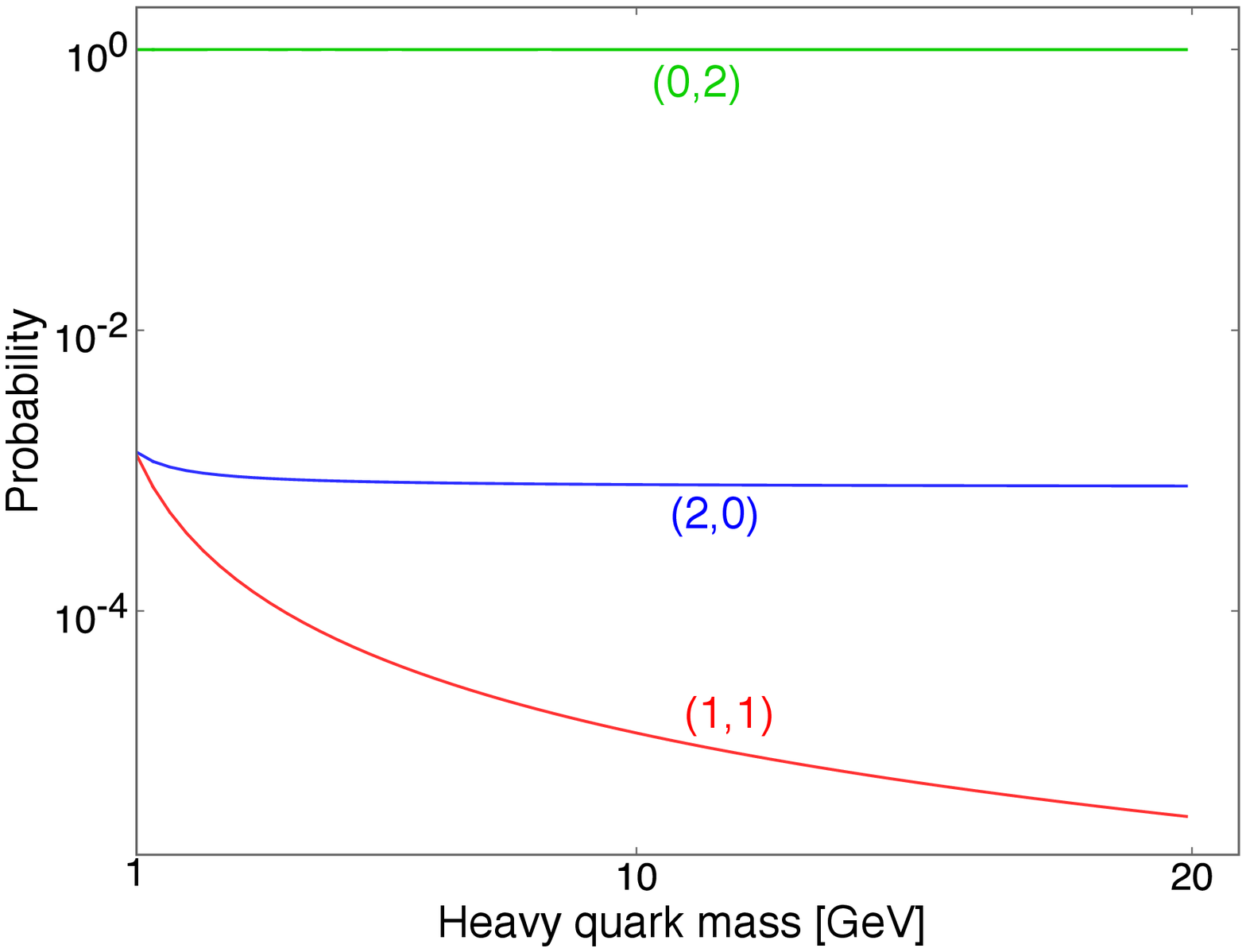}
        \end{center}
      \end{minipage}

    \end{tabular}
    \caption{The heavy quark mass dependences of the probabilities of (1,1) component (red line), (2,0) component (blue line) and (0,2) component (green line)  for $\Lambda(3/2_1^+)$.}
  \label{lambda32K1}
  \end{center}
\end{figure*}

\begin{figure*}[htbp]
  \begin{center}
    \begin{tabular}{c}

      \begin{minipage}{0.97\hsize}
        \begin{center}
          \includegraphics[trim = 0 370 0 70,clip, width=12.8cm]{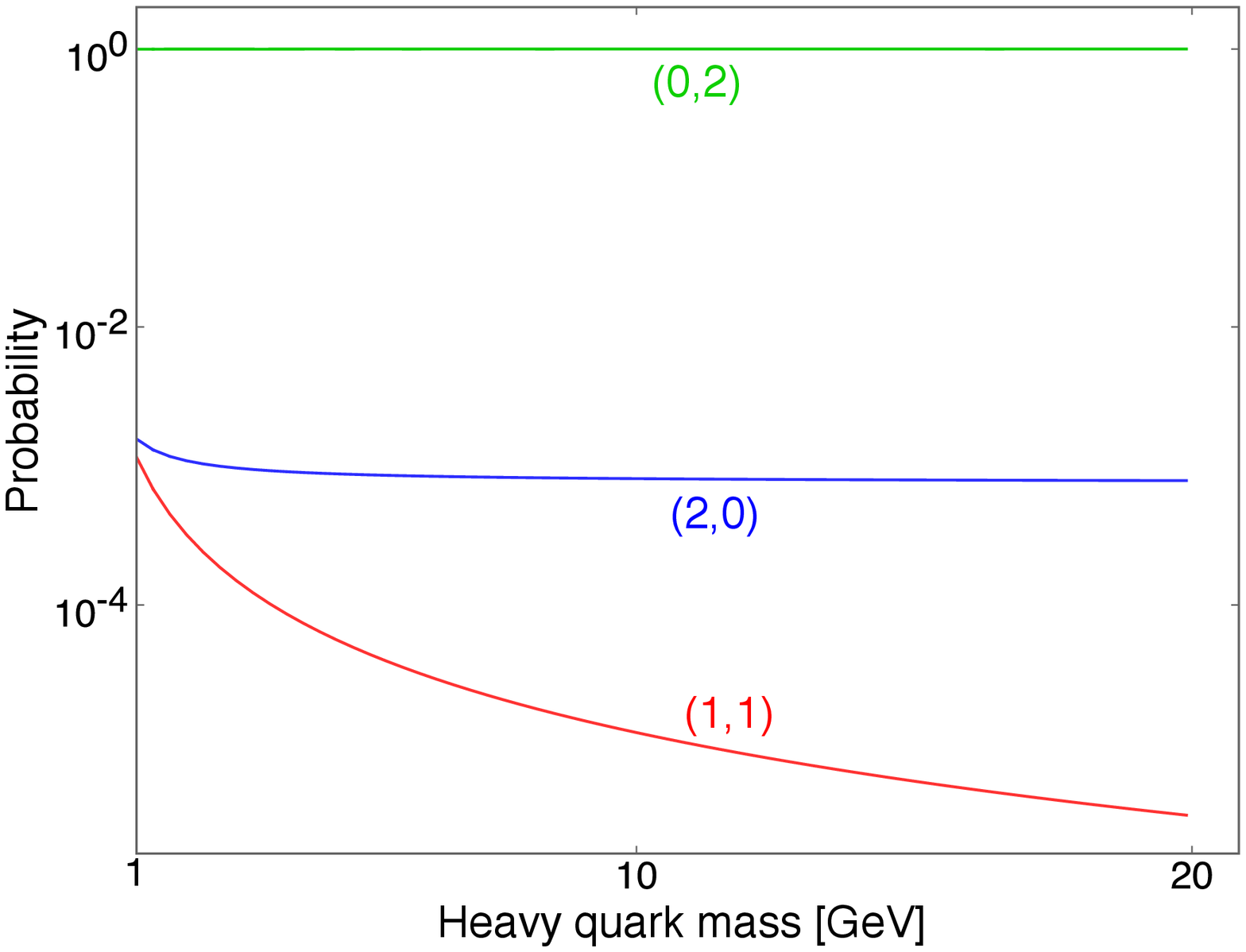}
        \end{center}
      \end{minipage}

    \end{tabular}
    \caption{The heavy quark mass dependences of the probabilities of (1,1) component (red line), (2,0) component (blue line) and (0,2) component (green line) for $\Lambda(5/2_1^+)$.}
      \label{lambda52K1}
  \end{center}
\end{figure*}

\begin{figure*}[htbp]
  \begin{center}
    \begin{tabular}{c}

      \begin{minipage}{0.97\hsize}
        \begin{center}
          \includegraphics[trim = 0 370 0 70,clip, width=12.8cm]{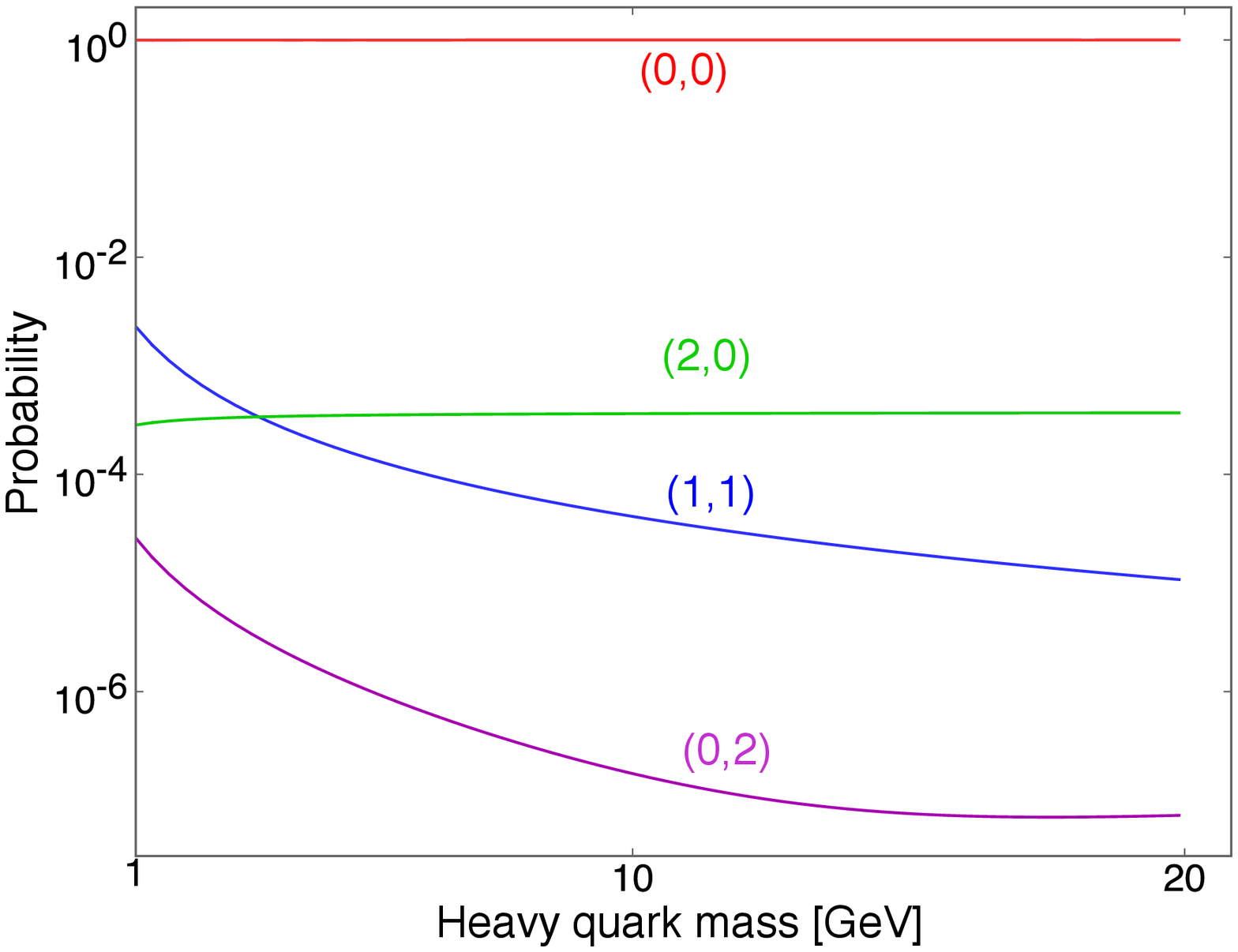}

        \end{center}
      \end{minipage}

    \end{tabular}
    \caption{The heavy quark mass dependences of the probabilities of the S-wave (0,0) component (red line), (1,1) component (blue line), (2,0) component (green line) and (0,2) component (violet line) for $\Sigma(1/2_1^+)$.}
      \label{sigma12K1}
  \end{center}
\end{figure*}

\begin{figure*}[htbp]
  \begin{center}
    \begin{tabular}{c}

      \begin{minipage}{0.97\hsize}
        \begin{center}
          \includegraphics[trim = 0 370 0 70,clip, width=12.8cm]{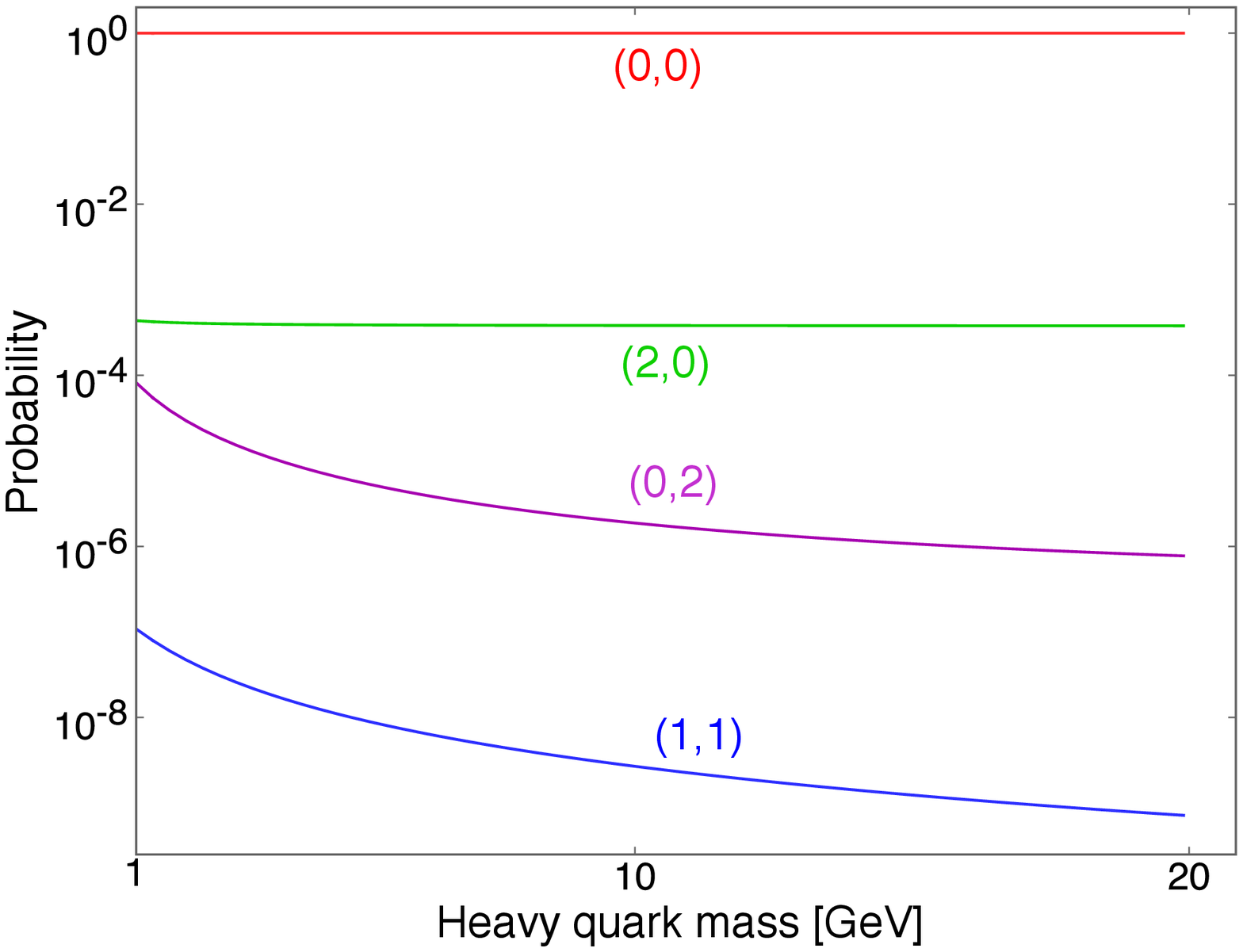}
        \end{center}
      \end{minipage}

    \end{tabular}
    
    \caption{The heavy quark mass dependences of the probabilities of the S-wave (l=0,L=0) component (red line), (1,1) component (blue line), (2,0) component (green line) and (0,2) component (violet line) for $\Sigma(3/2_1^+)$.}
      \label{sigma32K1}
  \end{center}
\end{figure*}
\parindent 15 pt

\newpage
\bibliography{ref.bib}

\end{document}